\documentclass[12pt]{article}
\usepackage{amssymb,amsmath,latexsym,graphicx, bm, mathrsfs, hyperref}
\usepackage{graphics,graphicx, cancel, ulem}
\usepackage[dvipsnames]{xcolor}

\usepackage[british]{datetime2}

\begin{document}

\title{Mechanics of geodesics in Information geometry \\ 
and Black Hole Thermodynamics  
\vspace{-0.45cm} 
\author{Sumanto Chanda$ ^*$, Tatsuaki Wada$ ^\dagger$ \\ 
\vspace{-0.45cm} \\ 
\textit{$ ^*$Indian Institute of Astrophysics} \\ 
\textit{Block 2, 100 Feet Road, Koramangala, Bengaluru,} 
\textit{Karnataka 560034, India.} \\ 
\texttt{\small $ ^*$\href{mailto:sumanto.chanda@iiap.res.in}{sumanto.chanda@iiap.res.in}} 
\smallskip \\ 
\textit{$ ^\dagger$Region of Electrical and Electronic Systems Engineering,} \\ 
\textit{Ibaraki University, Nakanarusawa-cho, Hitachi 316-8511, Japan} \\ 
\texttt{\small $ ^\dagger$\href{mailto:tatsuaki.wada.to@vc.ibaraki.ac.jp}
{tatsuaki.wada.to@vc.ibaraki.ac.jp}} }}

\maketitle

\thispagestyle{empty}
\vspace{-1.00cm}
\abstract{
In this article we shall discuss the theory of geodesics 
in information geometry, and an application in astrophysics. 
We will study how gradient flows in information geometry 
describe geodesics, 
explore the related mechanics by introducing a constraint, 
and apply our theory to Gaussian model and black hole 
thermodynamics.
Thus, we demonstrate how deformation of gradient flows leads 
to more general Randers-Finsler metrics, describe Hamiltonian 
mechanics that derive from a constraint, and prove duality 
via canonical transformation. 
We also verified our theories for a deformation of the Gaussian 
model, and described dynamical evolution of flat metrics for Kerr 
and Reissner-Nordstr\"om black holes.
}

\vspace{-0.5cm}
\tableofcontents

\numberwithin{equation}{section}

\setlength{\oddsidemargin}{-0.68in} % Left margin of 1 in + 0 in = 1 in
\setlength{\textwidth}{8.0in}   % Right margin of 8.5 in - 1 in - 6.5 in = 1 in
\setlength{\topmargin}{-0.75in}  % Top margin of 2 in -0.75 in = 1 in
\setlength{\textheight}{9.2in}  % Lower margin of 11 in - 9 in - 1 in = 1 in 

\newpage

\section{Introduction}
\setcounter{page}{1}

The origin of information geometry (IG) can be traced to Rao \cite{rao} 
who was the first to treat the Fisher matrix as a Riemannian metric, 
while the modern theory was largely the result of the efforts 
of Amari \cite{amari, amanag}. 
The subject classically treats a parametrized statistical model 
as a Riemannian manifold, using the Fisher information metric as a natural 
choice. 
However a modern treatment is based on the geometrical tools developed 
in affine differential geometry as described in \cite{amaribook}.
\smallskip

In the studies of gradient flows of IG by Fujiwara \cite{fujiwara}, 
Amari \cite{fujama}, and Nakamura \cite{nakamura}, it was shown 
that on a dually flat space $(\mathcal M_S, g, \nabla, \nabla^\star)$ 
with a potential function $U (p) := D_\nabla (p, q)$ at a fixed point $q$, 
the nonlinear gradient-flow equations in affine co-ordinates $\{ \theta \}$ 
are equivalent to linear differential equations in its dual affine co-ordinates 
$\{ \eta \}$. 
These linear differential equations 
\begin{equation} \label{lindiff} 
\frac{d \eta^g_i}{d t} = - \eta^g_i 
\quad , \quad i = 1, 2, . . . , 2m, 
\end{equation} 
are expressed as Hamilton's equations in spaces of even 
dimensions for their proposed Hamiltonian
\begin{equation} \label{hamprop} 
H^g = - \sum^m_{k = 1} Q^k P_k, 
\end{equation} 
where $Q^k$ and $P_k$ represent the generalized position 
and momentum variables respectively, and each individual term 
$Q^k P_k$ summed up in Eq. \eqref{hamprop} is a first integral 
of motion in its own right. 
\smallskip 

Boumuki and Noda studied the puzzling relation between conservative 
Hamiltonian flows $(Q^k (t), P_k (t))$ and dissipative gradient flows 
$\eta^g_i$ using symplectic geometry \cite{bn}. 
However, further physical interpretation was discussed alongside 
Scarfone and Matsuzoe in \cite{wsm1} from the view point of the analogies 
between the geodesic Hamiltonian mechanics related to Huygens' equations 
in geometric optics and the gradient-flow in IG \cite{wsm2}. 
Since the space of geometric optics is one of the Finsler spaces which are 
familiar in physics, we think it is important that the study of the gradient flows 
in IG involves Finsler geometry. 
\smallskip

Randers introduced his version of Finsler geometry in 1941 \cite{randers}, 
which involved a metric comparable to a Lagrangian describing mechanics 
of a point particle in curved space and electromagnetic gauge fields. 
This simple version of Finsler geometry generalizes Riemannian geometry 
by adding a linear term to the metric. 
Recently, the formulation of Jacobi metric in Randers-Finsler 
geometry have been studied in detail by Chanda, Guha, Gibbons, 
Maraner, and Werner in \cite{cggmw, cg, chanda}. 
The concept of a constraint for point particles in RF geometry 
was introduced in \cite{cggmw}, properly formalized and shown 
to act as a generator of equations of motion in \cite{cg, chanda} 
in a manner comparable to a Hamiltonian. This formulation of 
mechanics has been discussed in a parametrization invariant manner 
and the use of constraint as substitute is necessary because the 
Hamiltonian is found to be vanishing for RF metrics. 
\smallskip

One area for application of IG in physics is the topic of thermodynamic 
geometry \cite{wsm3}, also seen in recent work by Cafaro, Luongo, 
Mancini, Quevodo, Alsing, and Lupo on geometric \cite{clmq} and quantum 
\cite{acllmq} aspects of thermodynamics. 
Two forms of Riemannian geometry employed to study thermodynamics 
were introduced by Weinhold \cite{weinhold} and Ruppeiner \cite{ruppeiner} 
which are applicable as IG as seen in \cite{clmq}. 
In recent years, there has been plenty of progress on the study of the 
thermodynamics of blackholes via IG \cite{abp1, ap, abp2}. 
This presents an opportunity to describe such geometry for black holes 
as dynamical systems. 
\smallskip

In this article, we shall discuss aspects of geometry, mechanics, 
and application to astrophysics for IG related to gradient flows. 
We first review the gradient-flow equations of IG, fundamental aspects 
of mechanics, Randers-Finsler metrics, and optical metrics. 
For the geometric aspect, we shall attempt to formulate geodesics 
in IG from the gradient-flow equations discussed alongside Scarfone, 
and Matsuzoe \cite{wsm2}. 
For the mechanical aspect, we shall compare the gradient flows 
to Hamilton-Jacobi equations, and demonstrate using Hamiltonian 
mechanics derived from the constraint that we can confirm their 
dynamical equations. 
Furthermore, we shall demonstrate how the two types of gradient 
flows are dually related to each other via a canonical transformation. 
We next test our theory by applying our results to the Gaussian model. 
Finally, for the astrophysical applications we apply our theories 
to develop upon work by Aman, Bengtsson, and Pidokrajt 
\cite{abp1, ap, abp2} on flat metrics in IG of black hole thermodynamics 
to describe them as dynamical systems.

\section{Preliminaries} 
\label{sec:prelim} 

In this section, we will briefly review the fundamentals of IG \cite{amari} 
and the discussion on the gradient-flow equations by Wada, Scarfone, 
and Matsuzoe in their article \cite{wsm2}, and the constraint formulation 
of mechanics in Randers-Finsler spaces.

\numberwithin{equation}{subsection}

\subsection{Gradient flows of IG} 
\label{sec:gradflow} 
 
One of the great important concepts in IG is a dually-flat statistical 
manifold $(\mathcal M, g, \nabla, \nabla^\star)$, in which we have 
the dual potential functions $\Psi^\star (\bm \eta)$ and $\Psi (\bm \theta)$. 
Here $\nabla$ and $\nabla^\star$ are torsion-less dual affine connections. 
 \smallskip 
 
\noindent 
First we introduce the two sets of gradient-flow equations, 
respectively given by:
\begin{equation} \label{gradient1} 
     \frac{d \theta^i}{d t} = 
     g^{ij} (\bm \theta) 
     \frac{\partial \Psi}{\partial \theta^j} 
\quad , \quad 
     \frac{d \eta_i}{d t} = 
     - g^\star_{ij} (\bm \eta) 
     \frac{\partial \Psi^\star}{\partial \eta_j}.
\end{equation} 
The associated dual affine coordinates $\theta^i$ and $\eta_i$ 
are obtained from the dual potentials respectively as:
\begin{equation} \label{potgrad} 
     \eta_i = 
     \frac{\partial \Psi (\bm \theta)}{\partial \theta^i} 
\quad , \quad 
     \theta^i = 
     \frac{\partial \Psi^\star (\bm \eta)}{\partial \eta_i}, 
\end{equation} 
such that we can write \cite{fujama}: 
\begin{equation} \label{legendre} 
     \Psi (\bm\theta) + \Psi^\star (\bm\eta) = \eta_i \theta^i. 
\end{equation} 
In addition the metric satisfies the following relations.
\begin{equation} \label{metric} 
     g_{ij} (\bm \theta) = 
     g^\star_{ij} (\bm \eta) = 
     \frac{\partial \eta_i}{\partial \theta^j} = 
     \frac{\partial^2 \Psi (\bm \theta)}{\partial \theta^i \partial \theta^j} 
\quad , \quad 
     g^{ij} (\bm \theta) = 
     {g^\star}^{ij} (\bm \eta) = 
     \frac{\partial \theta^i}{\partial \eta_j} = 
     \frac{\partial^2 \Psi^\star (\bm \eta)}{\partial \eta^i \partial \eta^j} .
\end{equation} 
Then using Eq. \eqref{potgrad} and Eq. \eqref{metric}, we can write the two 
gradient-flow equations Eq. \eqref{gradient1} respectively as: 
\[ 
     \begin{split} 
          g_{ij} (\bm \theta) 
          \frac{d \theta^j}{d t} = 
          \frac{\partial \Psi}{\partial \theta^i} 
     \qquad &\Rightarrow \qquad 
          \frac{\partial^2 \Psi (\bm \theta)}{\partial \theta^i \partial \theta^j} 
          \frac{d \theta^j}{d t} = 
          \frac{d \ }{d t} 
          \left( 
               \frac{\partial \Psi}{\partial \theta^i} 
          \right) = 
          \frac{\partial \Psi}{\partial \theta^i},
     \\ 
          {g^\star}^{ij} (\bm \eta) 
          \frac{d \eta_i}{d t} = 
          - \frac{\partial \Psi^\star}{\partial \eta_j} 
     \qquad &\Rightarrow \qquad 
          \frac{\partial^2 \Psi^\star (\bm \eta)}{\partial \eta^i \partial \eta^j} 
          \frac{d \eta_j}{d t} = 
          \frac{d \ }{d t} 
          \left( 
               \frac{\partial \Psi^\star}{\partial \eta_i} 
          \right) = 
          - \frac{\partial \Psi^\star}{\partial \eta_i},
     \end{split} 
\]
\begin{equation} \label{gradient3} 
     \boxed{
          \frac{d \eta_i}{d t} = 
          \eta_i
     \quad , \quad 
          \frac{d \theta^i}{d t} = 
          - \theta^i.
     }
\end{equation} 
There exists a coordinate system in which the connection 
becomes zero. Such a coordinate system is called 
\textit{affine coordinate}.
The coefficients of $\nabla$ in the $\theta$-coordinate system 
and those of $\nabla^\star$  in the $\eta$-coordinate system 
respectively are: 
\begin{equation}
     \Gamma_{ijk}(\theta) = 0 
\quad , \quad 
     \Gamma^\star_{ijk}(\eta) = 0.
\end{equation}
Hence $\theta$ and $\eta$ coordinate systems 
are dual affine coordinates. 
For more details for the basic of IG, please refer 
to Amari's book \cite{amari}. 
\smallskip 

\noindent 
Now if we combine these two differential equations, 
we would obtain
$$
     \frac{d \eta_i}{d t} \theta^j + 
     \eta_i \frac{d \theta^j}{d t} = 
     \eta_i \theta^j + 
     \eta_i  (-\theta^j) = 0 
\quad \Rightarrow \quad 
     \frac{d \ }{d t} 
     \left( 
          \eta_i \, \theta^j 
     \right) = 0,
$$ 
\begin{equation} \label{indcons1} 
     \eta_i \, \theta^j = const. 
\end{equation} 
Thus, according to Eq. \eqref{indcons1}, each and every one 
of the terms of the inner product: 
$$\sum_i \eta_i \,\theta^i = const.,$$ 
would describe a constant of motion in their own individual right. 
This is obvious since, if we solve both equations of Eq. \eqref{gradient3} 
individually, we will have: 
\begin{equation} \label{gradsol} 
     \eta_i (t) = 
     \eta_i (0) 
     \exp (t) 
\quad , \quad 
     \theta^i (t) = 
     \theta^i (0) 
     \exp (- t), 
\end{equation} 
where $\eta_i (0)$, $\theta^i (0)$ are constants. 
This means that Eq. \eqref{indcons1} will become: 
\begin{equation} \label{indcons2} 
     \eta_i \, \theta^j = 
     \eta_i (0) \, \theta^j (0) := K_i^j \quad
     (\textrm{constant}). 
\end{equation}
Under these circumstances, using Eq. \eqref{gradient3} 
and Eq. \eqref{indcons2} we shall have the potential 
$\Psi (\bm \theta)$ given by: 
\begin{equation} \label{indpot} 
     \Psi (\bm \theta) = 
     \int d \theta^i \eta_i = 
     - \int d t \; \theta^i \eta_i = 
     - t \; \text{Tr} \left( K \right) = 
     \ln \left( \frac{\theta^j (t)}{\theta^j (0)} \right) \; 
     \text{Tr} \left( K \right), 
\end{equation} 
and the metric according to Eq. \eqref{metric} would be 
(repeated indices do not mean summation in the following equation): 
\begin{equation} \label{indmetric} 
     g_{ij} (\bm \theta) = 
     \frac{\partial \eta_i}{\partial \theta^j} = 
     - K^m_i 
     \frac1{\left( \theta^m \right)^2} 
     \delta^m_j = 
     - K^j_i 
     \frac1{\left( \theta^j \right)^2}. 
\end{equation} 
Note that the Hamiltonian $H^g$ in Eq. \eqref{hamprop} 
has the same structure.
However, we emphasize that the two differential equations 
of Eq. \eqref{gradient3} describe different processes in general 
and not necessarily combine them at all. 
\smallskip 

Finally, we focus on the parameter $t$ in the gradient-flow 
equations Eq. \eqref{potgrad}, which is a non-affine parameter.
It is known that if a geodesic curve, say $x^i = x^i(s)$, 
is written by the geodesic equation
\begin{equation}
     \frac{d^2 x^i(s)}{d s^2} + 
     \Gamma^i_{j k}(\bm{x})  
     \frac{d x^j (s)}{d s}  
     \frac{d x^k (s)}{d s} = 0,
\end{equation}
the parameter $s$ is called an affine parameter. 
From the first differential equation in Eq. \eqref{gradient3}, 
we have
\begin{equation}
     \frac{d^2 \theta^i}{d t^2} = - \frac{d \theta^i}{dt},
\end{equation}
which is the geodesic equation in the $\theta$-coordinate 
system because $\Gamma^i_{jk}(\bm{\theta}) = 0$.
Hence the parameter $t$ is a non-affine parameter.

\subsection{Aspects of Classical Mechanics} 
\label{sec:cmrf} 

Although Maupertuis is credited for introducing the principle of least 
action, he had applied the principle only to light \cite{maup}. 
It was evidently Euler who had intuitively connected the principle 
to mechanics \cite{euler}. 
Here, we will review the fundamentals from classical mechanics 
relevant to our discussion on studying IG using a Lagrangian mechanics 
approach later on. 
This involves reviewing principle of least action via Euler-Lagrange 
equation and Maupertuis principle, and canonical transformation. 
\bigskip 

\noindent 
If we describe a curve parametrized wrt $\tau$ and an action integral 
along it with a Lagrangian $L (\bm{x, \dot x})$ 
\begin{equation} \label{act1} 
     S = \int_1^2 d \tau \; L (\bm{x, \dot x}) 
\quad , \quad \text{where } \ 
     \bm{\dot x} = \frac{d {\bm x}}{d \tau},
\end{equation} 
then upon variation of action Eq. \eqref{act1}, we will have: 
\begin{equation} \label{var} 
     \therefore \qquad 
     \delta S = 
     \int_1^2 d \tau 
     \left[ 
          \frac{\partial L}{\partial x^i} - 
          \frac{d \ }{d \tau} 
          \left( 
               \frac{\partial L}{\partial \dot x^i} 
          \right) 
     \right] 
     \delta x^i + 
     \left[ 
          \left( 
               \frac{\partial L}{\partial \dot x^i} 
               \delta x^i 
          \right) 
     \right]_1^2. 
\end{equation} 
Assuming that $\delta x^i = 0$ at end points, 
for the dynamical path described by minimum value 
of action Eq. \eqref{act1}, the action variation Eq. \eqref{var} 
leads to the Euler-Lagrange equation: 
\begin{equation} \label{eulag1} 
     \delta S = 0 
\quad \Rightarrow \quad 
     \frac{\partial L}{\partial x^i} - 
     \frac{d \ }{d \tau} 
     \left( 
          \frac{\partial L}{\partial \dot x^i} 
     \right) = 0. 
\end{equation} 
If we define the ``momentum'' $\bm p$ here as: 
\begin{equation} \label{canmom} 
     p_i = \frac{\partial L}{\partial \dot x^i}, 
\end{equation} 
then along the classical path of least action, we can describe 
the Maupertuis principle based on how the action varies at the 
end points: 
\begin{equation} \label{maup} 
     \delta S = 
     \left[ 
          p_i \; \delta x^i 
     \right]_1^2 
\quad \Rightarrow \quad 
     \frac{\partial S}{\partial x^i} = p_i 
\quad \Rightarrow \quad 
     \boxed{
          d S = p_i \; d x^i
     }. 
\end{equation}
Furthermore, if we take the derivative of the Lagrangian 
along the curve and apply Eq. \eqref{eulag1}, we will have: 
$$
     \frac{d L}{d \tau} = 
     \frac{\partial L}{\partial x^i} 
     \dot x^i + 
     \frac{\partial L}{\partial \dot x^i} 
     \ddot x^i = 
     \underbrace{ 
          \left[ 
               \frac{\partial L}{\partial x^j} - 
               \frac{d \ }{d \tau} 
               \left( 
                    \frac{\partial L}{\partial \dot x^i} 
               \right) 
          \right] 
     }_0 
     \dot x^i + 
     \frac{d \ }{d \tau} 
     \left( 
          \frac{\partial L}{\partial \dot x^i} 
          \dot x^i 
     \right)
$$ 
\begin{equation} \label{ham} 
     \Rightarrow \qquad 
     \frac{d H}{d \tau} = 
     \frac{d \ }{d \tau} 
     \left( 
          p_i \dot x^i - L 
     \right) = 0 
\quad \Rightarrow \quad 
     \boxed{H = p_i \dot x^i - L = const.}.
\end{equation} 
Here, if we perform a canonical transformation \cite{goldstein}: 
\begin{equation} \label{can} 
     G (\bm{x, x^\star}): \ 
     \left( x^i, p_i \right) 
\longrightarrow 
     \left( {x^\star}^i, p^\star_i \right), 
\end{equation} 
then we can formulate a new Lagrangian $L^\star$ by adding 
a total time derivative without affecting the equations of motion: 
\begin{equation} \label{infocan} 
     L^\star = L - \frac{d G}{d \tau},
\end{equation} 
such that the Maupertuis form of the Lagrangian Eq. \eqref{maup} 
according to Eq. \eqref{infocan} changes as: 
$$
     p^\star_i \dot x^{\star i} = 
     p_i \dot x^i - 
     \frac{\partial G}{\partial x^i} 
     \dot x^i - 
     \frac{\partial G}{\partial {x^\star}^i} 
     \dot x^{\star i}
$$
\begin{equation} \label{caneq} 
     \therefore \qquad 
     p^\star_i = 
     - \frac{\partial G}{\partial {x^\star}^i} 
\quad , \quad 
     p_i = 
     \frac{\partial G}{\partial x^i} . 
\end{equation} 
If we now perform the following canonical transformation 
\cite{goldstein}: 
\begin{equation} \label{canonical} 
     G (\bm{x, p}) = 
     p_i x^i: \ 
     \left( x^i, p_i \right) 
\longrightarrow 
     \left( 
          p_i, - x^i 
     \right),
\end{equation} 
where clearly $\bm{x^\star} = \bm{p}$, according 
to Eq. \eqref{caneq}, we can write: 
\begin{equation} \label{canspec} 
     {x^\star}^i = p_i
\quad , \quad 
     p^\star_i = 
     - \frac{\partial G}{\partial p_i} = 
     - x^i, 
\end{equation} 
according to which, we can rewrite the Maupertuis form 
of the canonical momentum in Eq. \eqref{maup} as: 
\begin{equation} \label{canmaup} 
     p_i = \frac{\partial S}{\partial x^i} 
\qquad \longrightarrow \qquad 
     {x^\star}^i = - \frac{\partial S}{\partial p^\star_i}
\end{equation} 
Thus, this transformation allows us to effectively 
swap the canonical variables in phase space, while 
preserving the dynamics and the form of the action.

\subsection{Geodesics in Randers-Finsler spaces} 
\label{sec:rfcons}

Finsler spacetimes were studied by Finsler in his dissertation in 
1918 and named after him in 1933 by Cartan. They are a more 
general form of spacetime beyond the Riemannian restriction 
essentially characterised by the following rules: 

\begin{enumerate} 

\item Triangle inequality 
$F (\bm x, \bm y + \bm z) \leq F (\bm x, \bm y) + F (\bm x, \bm z) 
\quad \forall \quad \bm y, \bm z \in T_{\bm x} M.$ 

\item Linearity
$F (\bm x, \Lambda \bm y) = \Lambda F (\bm x, \bm y) 
\quad \forall \quad \Lambda \geq 0 \ 
(\text{not necessarily for } \Lambda < 0).$ 

\item Posetive definiteness 
$F (\bm x, \bm y) > 0 \quad \forall \quad \bm y \neq 0.$

\end{enumerate}

\noindent
In 1941, Randers \cite{randers} modified a Riemannian metric 
$g = g_{ij} (\bm x) \ dx^i \otimes dx^j$ into a Finsler metric by 
adding a linear term $A = A_i (\bm x) dx^i$. This form simultaneously 
accounts for the influence of curvature and gauge 
fields in point particle motion. The resulting Randers-Finsler (RF)
metric on the tangent space $T_x M$ is given by:
$$
     F (\bm x, \bm y) = 
     \sqrt{g_{ij} (\bm x) y^i y^j} + 
     A_i (\bm x) y^i , 
\qquad 
     y = y^i \partial_i \in T_x M.
$$ 

\noindent
Mechanics in RF geometry \cite{randers, chanda} is analogous 
to a particle in gauge fields, where the first part of the metric with 
the square root of the norm accounts for the influence of curvature 
in the action, while the linear term outside accounts for gauge field 
interaction. 
In RF geometry, the second part is geometric in origin. 

\bigskip 

If the worldline length $s$ of a curve between two points given 
by integration of the metric $d s$ is parametrised by $\tau$, 
it can be written in terms of a Lagrangian $L$ like an action 
integral Eq. \eqref{act1}:
\begin{equation} \label{action} 
     s = \int_1^2 d s = 
     \int_1^2 d \tau \ L (\bm x, \bm{\dot x}) 
\quad , \quad 
     \text{where } \ L = \frac{d s}{d \tau}.
\end{equation} 

\noindent
Consider the RF metric given by 
\begin{equation} \label{rander} 
     ds = 
     \sqrt{
          g_{ij} (\bm x) 
          d x^i d x^j
     } + 
     A_i (\bm x) d x^i. 
\end{equation} 
Based on Eq. \eqref{action} we can deduce the Lagrangian from Eq. \eqref{rander}, 
and the momentum according to Eq. \eqref{canmom}: 
$$
     L = 
     \sqrt{
          g_{ij} (\bm x) 
          \dot x^i \dot x^j
     } + 
     A_i (\bm x) \dot x^i
$$
\begin{equation} \label{rfmom} 
     p_i = 
     \frac{\partial L}{\partial \dot x^i} = 
     g_{ij} (\bm x) 
     \frac{d x^j}{d \lambda} + 
     A_i (\bm x), 
\qquad \text{where } \ 
     d \lambda = 
     \sqrt{g_{ij} (\bm x) \dot x^i \dot x^j}
\end{equation} 
Motion will occur along the curve described by the solution of the 
Euler-Lagrange equation Eq. \eqref{eulag1}, called the geodesic along 
which the metric is given by the Maupertuis principle Eq. \eqref{maup}:
\begin{equation} \label{randmaup} 
     d s = 
     \frac{\partial s}{\partial x^i} 
     d x^i = 
     p_i d x^i
\quad \Rightarrow \quad 
     p_i = 
     \frac{\partial s}{\partial x^i}.
\end{equation} 
Using the Lagrangian $L$ defined according to Eq. \eqref{action}, 
the canonical momenta $\bm p$ according to Eq. \eqref{rfmom} 
leads us to the gauge-covariant momenta $\bm \pi$ given by: 
\begin{equation} \label{gcmom} 
     \pi_i = p_i - A_i (\bm x) = 
     \frac{g_{ij} (\bm x) \dot x^j}
     {
          \sqrt{
               g_{\alpha \beta} (\bm x) 
               \dot x^\alpha \dot x^\beta
          }
     } = 
     g_{ij} (\bm x) 
     \frac{d x^j}{d \lambda}, 
\qquad \text{where } \ 
     d \lambda := 
     \sqrt{
          g_{\alpha \beta} (\bm x) 
          d x^\alpha d x^\beta
     }.
\end{equation} 
The gauge-covariant momenta $\bm \pi$ Eq. \eqref{gcmom} obey 
the constraint:
\begin{equation} \label{cons1} 
     \phi (\bm x, \bm p) = 
     \sqrt{g^{ij} (\bm x) \pi_i \pi_j} = 
     \sqrt{g_{ij} (\bm x) 
     \frac{d x^i}{d \lambda} 
     \frac{d x^j}{d \lambda}} = 1,
\end{equation} 
which acts as a generator of equations of motion, the constraint 
equivalent of Hamilton's equations of motion:
\begin{equation} \label{conseq3} 
     \boxed{
          \frac{d x^i}{d \lambda} = 
          \frac{\partial \phi}{\partial p_i} 
     \quad , \quad 
          \frac{d p_i}{d \lambda} = 
          - \frac{\partial \phi}{\partial x^i}}.
\end{equation} 
Although the RF Lagrangian exactly matching the Maupertuis form 
prevents us from deducing a Hamiltonian function, it also allows us 
to determine the metric using the Maupertuis principle. 
When starting from the constraint, this is done by applying Eq. \eqref{rfmom} 
and Eq. \eqref{conseq3} to Maupertuis principle Eq. \eqref{randmaup}:
\begin{equation} \label{maup2} 
     d s = p_i d x^i = 
     g_{ij} (\bm x) 
     \frac{\partial \phi}{\partial p_j} d x^i + 
     A_i (\bm x) d x^i = 
     d \lambda + 
     A_i (\bm x) d x^i.
\end{equation}
From Eq. \eqref{cons1}, we can write 
$d \lambda = \sqrt{g_{ij} (\bm x) d x^i d x^j}$,
which upon application to Eq. \eqref{maup2}, will give us the original 
RF metric Eq. \eqref{rander}.

\subsection{Geodesics of optical metrics} 
\label{sec:aniso}

Consider the stationary spacetime metric given below where we set $c = 1$: 
\begin{equation} \label{stat} 
     ds^2 = 
     g_{00} (\bm x) dt^2 + 
     2 g_{i0} (\bm x) \ d t \ dx^i + 
     g_{ij} (\bm x) dx^i dx^j.
\end{equation} 
From the null version of Eq. \eqref{stat} we write the Optical metric $d s_\mathcal O$ 
as a solution of the quadratic equation for $dt$ 
\begin{equation} \label{optical} 
     d s_{\mathcal O} = dt = 
     \pm \sqrt{
          - \frac{\gamma_{ij} (\bm x)}{g_{00} (\bm x)} d x^i d x^j
     } - 
     \frac{g_{i0} (\bm x)}{g_{00} (\bm x)} dx^i, 
\quad 
     \text{where } \ 
     \gamma_{ij} (\bm x) = g_{ij} (\bm x) - 
     \frac{g_{i0} (\bm x) g_{j0} (\bm x)}{g_{00} (\bm x)},
\end{equation} 
which we can see is a Randers type of Finsler metric \cite{randers}, 
where we will take $+$ solution for $dt > 0$. 
If we define the conformal refractive index factor $n (\bm x)$ as: 
\begin{equation} \label{confri} 
     n (\bm x) = \frac1{\sqrt{- g_{00} (\bm x)}} 
\qquad , \quad 
     \text{where } \ 
     g_{00} (\bm x) < 0,
\end{equation} 
then the RF metric Eq. \eqref{optical} can be written as: 
\begin{equation} \label{conform} 
d s_{\mathcal O} = 
     \sqrt{
          \left( n (\bm x) \right)^2 
          \gamma_{ij} (\bm x) d x^i d x^j
     } + 
     \alpha_i (\bm x) d x^i, 
\quad 
     \text{where } \ 
     \alpha_i (\bm x) = 
     \left( n (\bm x) \right)^2 
     g_{i0} (\bm x), 
\end{equation} 
and the gauge-covariant momenta are given by: 
\begin{equation} \label{optmom} 
     \Pi_i = p_i - \alpha_i (\bm x) = 
     \left( n (\bm x) \right)^2 
     \gamma_{ij} (\bm x) \frac{d x^i}{d \lambda}, 
\quad 
     \text{where } \ 
     d \lambda = 
     \sqrt{
          \left( n (\bm x) \right)^2 
          \gamma_{ij} (\bm x) d x^i d x^j
     }. 
\end{equation} 
The constraint for this optical metric is given by: 
\begin{equation} \label{optcons} 
     \phi (\bm{x, p}) = 
     \sqrt{
          \frac1{\left( n (\bm x) \right)^2}g^{ij} (\bm x) 
          \Pi_i \Pi_j
     } = 1, 
\quad 
     \text{where } \ 
     g^{ij} (\bm x) 
     \gamma_{jk} (\bm x) = 
     \delta^i_k. 
\end{equation} 
Based on the equation: 
$$
     \frac{d \phi}{d \lambda} = 
     \frac{\partial \phi}{\partial x^i} 
     \frac{d x^i}{d \lambda} + 
     \frac{\partial \phi}{\partial p_i} 
     \frac{d p_i}{d \lambda} = 0,
$$
the constraint equations Eq. \eqref{conseq3} upon applying Eq. \eqref{optmom} 
are given by: 
\begin{equation} \label{conseq1} 
     \begin{split}
          \frac{d x^i}{d \lambda} &= 
          \frac{\partial \phi}{\partial p_i} = 
          \frac1{\left(n (\bm x) \right)^2} g^{ij} (\bm x) \Pi_j 
     \\ 
          \frac{d p_i}{d \lambda} &= 
          - \frac{\partial \phi}{\partial x^i} = 
          - \frac1{n (\bm x)} 
          \left[ 
               \frac{\partial \ }{\partial x^i} 
               \left( 
                    \sqrt{
                         g^{ab} (\bm x) \Pi_a \Pi_b
                    } 
               \right) - 
               \frac{\partial n}{\partial x^i} 
          \right]. 
     \end{split}
\end{equation}

\numberwithin{equation}{section}

\section{Geodesics in IG} 
\label{sec:iggeodesic} 

Previously in Subsec. \ref{sec:cmrf}, we reviewed the general 
formulation of Lagrangian mechanics, and in Subsec. \ref{sec:rfcons} 
its specific application to RF spaces. 
This time, we shall formulate a Riemannian metric with a conformal 
factor in IG starting from the gradient-flow equations. 
\bigskip  

First, we shall demonstrate the similarity between mechanics 
in Riemannian geometry and dynamics in IG. 
If we write if $A_i (\bm x) = 0$ for Riemannian metric 
in Eq. \eqref{rander}, then we will have: 
\begin{equation} \label{riem} 
     ds = 
     \sqrt{
          g_{ij} (\bm x) 
          d x^i d x^j
     }. 
\end{equation} 
Based on Eq. \eqref{action} we can deduce the Lagrangian from 
Eq. \eqref{riem}, and subsequently derive the momentum: 
\begin{equation} \label{riemmom} 
     L = 
     \sqrt{
          g_{ij} (\bm x) 
          \dot x^i \dot x^j
     }
\quad \Rightarrow \quad 
     p_i = 
     \frac{\partial L}{\partial \dot x^i} = 
     g_{ij} (\bm x) 
     \frac{d x^j}{d s}. 
\end{equation} 
Thus, according to Eq. \eqref{randmaup} we can write from 
the momentum Eq. \eqref{riemmom} that 
\begin{equation} \label{hamjacobi1} 
     p_i = 
     g_{ij} (\bm x) 
     \frac{d x^j}{d s} = 
     \frac{\partial s}{\partial x^i}
\quad \Rightarrow \quad 
     \frac{d x^i}{d s} = 
     g^{ij} (\bm x) 
     \frac{\partial s}{\partial x^j}. 
\end{equation} 
Here, if we identify a conformal factor within the metric tensor 
in Eq. \eqref{riem} such that 
$g_{ij} (\bm x) = \left[ C (\bm x) \right]^{-2} h_{ij} (\bm x)$, 
then Eq. \eqref{hamjacobi1} shall become: 
\begin{equation} \label{hamjacobi2} 
     g^{ij} (\bm x) = 
     \left[ C (\bm x) \right]^2 
     h^{ij} (\bm x)
\quad \Rightarrow \quad 
     \frac{d x^i}{d t} = 
     h^{ij} (\bm x) 
     \frac{\partial s}{\partial x^j}, 
\quad \text{where } \ 
     d t = 
     \left[ C (\bm x) \right]^2 
     d s, 
\end{equation} 
which is comparable to the first gradient-flow equation 
in Eq. \eqref{gradient1}. 
Furthermore, the metric can be written according to the Maupertuis 
principle Eq. \eqref{randmaup}.

\numberwithin{equation}{subsection}

\subsection{Riemannian metric from gradient flows} 
\label{sec:inforiem} 

\subsubsection*{From the first gradient flow} 

If we start with the first of the two gradient-flow equations from 
Eq. \eqref{gradient1}, then we can write: 
\begin{equation} \label{norm1}  
     \frac{d \theta^i}{d t} = 
     g^{ij} (\bm \theta) 
     \frac{\partial \Psi}{\partial \theta^j} 
\quad \Rightarrow \quad 
     g_{ij} (\bm \theta) 
     \frac{d \theta^i}{d t} 
     \frac{d \theta^j}{d t} = 
     g^{ij} (\bm \theta) 
     \frac{\partial \Psi}{\partial \theta^i} 
     \frac{\partial \Psi}{\partial \theta^j} = 
     \left[ C (\bm \theta) \right]^2,
\end{equation} 
where $\left[ C (\bm \theta) \right]^{-1}$ is comparable 
to the conformal refractive index factor discussed 
in Eq. \eqref{confri}. 
We can introduce the conformal Riemannian metric
\begin{equation}
 G_{ij}(\bm \theta) := 
 \left[ C (\bm \theta) \right]^{-2} 
 g_{ij} (\bm \theta),
 \label{Gij}
 \end{equation}
 and solve for $dt$ from Eq. \eqref{norm1} as
\begin{equation} \label{igriem1} 
     d t = 
     \sqrt{ 
          \left[ C (\bm \theta) \right]^{-2} 
          g_{ij} (\bm \theta) 
          d \theta^i d \theta^j 
     } = 
     \sqrt{ 
          G_{ij} (\bm \theta) 
          d \theta^i d \theta^j 
     }.
\end{equation} 
which is comparable to the optical metric in Eq. \eqref{conform}. 
Furthermore, using Eq. \eqref{igriem1} and the first gradient flow 
equation Eq. \eqref{gradient1}, we can write: 
\begin{equation} \label{igaction1} 
     d \Psi = 
     \frac{\partial \Psi}{\partial \theta^j} d \theta^i = 
     g_{ij} (\bm \theta) 
     \frac{d \theta^j}{d t} 
     d \theta^i = 
     \sqrt{ 
          \left[ C (\bm \theta) \right]^2 
          g_{ij} (\bm \theta) 
          d \theta^i d \theta^j 
     } = 
     \left[ C (\bm \theta) \right]^2 
     d t.
\end{equation} 
The null metric we can determine from the above optical 
metric Eq. \eqref{igriem1} is: 
\begin{equation} \label{ignull1} 
     d s^2 = 
          g_{ij} (\bm \theta) d \theta^i d \theta^j - 
          \left[ C (\bm \theta) \right]^2 
          d t^2 
     = 0.
\end{equation} 
If we define a Lagrangian from the metric Eq. \eqref{igriem1} 
according to Eq. \eqref{action}, then we have: 
\begin{equation} \label{iglag1} 
     L({\bm \theta}, \dot{ \bm \theta})  = \frac{d t}{d \tau} = 
     \sqrt{ 
          \left[ C (\bm \theta) \right]^{-2} 
          g_{ij} (\bm \theta) 
          \dot \theta^i \dot \theta^j 
     } = 
     \sqrt{ 
          G_{ij} (\bm \theta) 
          \dot \theta^i \dot \theta^j 
     }, 
\end{equation} 
then according to Eq. \eqref{potgrad} and Eq. \eqref{gradient1}, 
the canonical momentum $p_i$ is given by: 
$$
     \eta_i = \frac{\partial \Psi}{\partial \theta^i} 
\quad \Rightarrow \quad 
     \frac{d \theta^i}{d t} = 
     g^{ij} (\bm \theta) 
     \frac{\partial \Psi}{\partial \theta^j} = 
     g^{ij} (\bm \theta) \eta_j,
$$
\begin{equation} \label{igmom1} 
     p_i := 
     \frac{\partial L}{\partial \dot \theta^i} = 
          \left[ C(\bm \theta) \right]^{-2} 
          g_{ij} (\bm \theta) 
          \frac{d \theta^j}{d t} = 
          \left[ C (\bm \theta) \right]^{-2} 
          \frac{\partial \Psi}{\partial \theta^i} = 
          \left[ C (\bm \theta) \right]^{-2} \eta_i, 
\end{equation} 
and by Maupertuis principle Eq. \eqref{maup}, we can write 
the Maupertuis form of the metric Eq. \eqref{igriem1} as: 
\begin{equation} \label{igmaup1} 
     d t = 
     p_i \; d \theta^i = 
     \left[ C(\bm \theta) \right]^{-2} \eta_i \; d \theta^i. 
\end{equation} 
Furthermore, if we apply Eq. \eqref{potgrad} and Eq. \eqref{igmom1} 
to Eq. \eqref{norm1}, then we can formulate a constraint: 
$$
     \frac1{\left[ C (\bm \theta) \right]^2} 
     g_{ij} (\bm \theta) 
     \frac{d \theta^i}{d t} 
     \frac{d \theta^j}{d t} = 
     \frac1{\left[ C (\bm \theta) \right]^2} 
     g^{ij} (\bm \theta) 
     \eta_i \eta_j = 1.
$$
\begin{equation} \label{igham1} 
     \Phi (\bm{\eta, \theta}) = 
     \sqrt{
          G^{ij} (\bm \theta) p_i p_j
     } = 
     \sqrt{
          \left[ C (\bm \theta) \right]^2 
          g^{ij} (\bm \theta) p_i p_j
     } = 
     \sqrt{
          \frac1{\left[ C (\bm \theta) \right]^2} 
          g^{ij} (\bm \theta) \eta_i \eta_j
     } = 1, 
\end{equation} 
\begin{equation} \label{igcons1} 
     \frac{d \Phi}{d t} = 
     \frac{\partial \Phi}{\partial \theta^i} 
     \frac{d \theta^i}{d t} + 
     \frac{\partial \Phi}{\partial \eta_i} 
     \frac{d \eta_i}{d t} = 0
\quad \Rightarrow \quad 
     \frac{\partial \Phi}{\partial \theta^i} 
     \frac{d \theta^i}{d t} = 
     - \frac1{\left[ C (\bm \theta) \right]^2} 
     \frac{d \theta^i}{d t}
     \frac{d \eta_i}{d t},
\end{equation} 
that generates the following constraint equations: 
\begin{equation} \label{igconseq11} 
     \frac{\partial \Phi}{\partial \eta_i} = 
     \frac1{\left[ C (\bm \theta) \right]^2} 
     g^{ij} (\bm \theta) 
     \eta_j = 
     \frac1{\left[ C (\bm \theta) \right]^2} 
     \frac{d \theta^i}{d t} = 
     \frac{d \theta^i}{d \lambda}, 
\qquad \text{where } \ 
     d \lambda := 
     \left[ C (\bm \theta) \right]^2 d t
\end{equation} 
\begin{equation} \label{igconseq12} 
     \frac{d \eta_i}{d \lambda} = 
     \frac1{\left[ C (\bm \theta) \right]^2} 
     \frac{d \eta_i}{d t} = 
     - \frac{\partial \Phi}{\partial \theta^i} = 
     - \frac12 \frac{\partial \ }{\partial \theta^i} 
     \left( 
          \frac{g^{jk} (\bm \theta) }{\left[ C (\bm \theta) \right]^2} 
     \right) 
     \eta_j \eta_k. 
\end{equation} 
Furthermore, by applying the metric rule Eq. \eqref{metric}, 
if we treat $\bm \eta$ as a function of $\bm \theta$ in 
Eq. \eqref{igcons1} and apply Eq. \eqref{igconseq11} and 
Eq. \eqref{igconseq12}, we can also write: 
$$
     \left( 
          \frac{\partial \Phi}{\partial \theta^i} + 
          \frac{\partial \Phi}{\partial \eta_j} 
          \frac{\partial \eta_j}{\partial \theta^i} 
     \right) 
     \frac{d \theta^i}{d t} = 0 
\quad \Rightarrow \quad 
     \frac{\partial \Phi}{\partial \theta^i} = 
     - g_{ij} (\bm \theta) 
     \frac{\partial \Phi}{\partial \eta_j} = 
     - \frac1{\left[ C (\bm \theta) \right]^2} 
     \eta_i, 
$$ 
\begin{equation} \label{rule1} 
     \Rightarrow \qquad 
     - \frac12 \frac{\partial \ }{\partial \theta^i} 
     \left( 
          \frac{g^{jk} (\bm \theta) }{\left[ C (\bm \theta) \right]^2} 
     \right) 
     \eta_j \eta_k = 
     \frac1{\left[ C (\bm \theta) \right]^2} \,
     \eta_i ,
\end{equation}  
\begin{equation} \label{iggradeq1} 
     \therefore \qquad 
     \frac{d \eta_i}{d \lambda} = 
     \frac1{\left[ C (\bm \theta) \right]^2} 
     \frac{d \eta_i}{d t} = 
     - \frac{\partial \Phi}{\partial \theta^i} = 
     \frac1{\left[ C (\bm \theta) \right]^2} 
     \eta_i 
\quad \Rightarrow \quad 
     \boxed{\frac{d \eta_i}{dt} = \eta_i,}
\end{equation}
which is the first equations of Eq. \eqref{gradient3}. 
In this way we have related the first gradient-flow 
to the geodesic flow described by the Lagrangian 
$L({\bm \theta}, \dot{ \bm \theta})$ 
in Eq. \eqref{igriem1} with the conformal Riemannian metric 
$G_{ij}(\bm \theta)$ 
in Eq. \eqref{Gij}.

\subsubsection*{From the second gradient flow}  

On the other hand, if we start with the second of the two gradient-flow equations 
from Eq. \eqref{gradient1}, then we can write: 
\begin{equation} \label{norm2} 
     \frac{d \eta_i}{d t^\star} = 
     - g^\star_{ij} (\bm \eta) 
     \frac{\partial \Psi^\star}{\partial \eta_j} 
\quad \Rightarrow \quad 
     {g^\star}^{ij} (\bm \eta) 
     \frac{d \eta_i}{d t^\star} 
     \frac{d \eta_j}{d t^\star} = 
     g^\star_{ij} (\bm \eta) 
     \frac{\partial \Psi^\star}{\partial \eta_i} 
     \frac{\partial \Psi^\star}{\partial \eta_j} = 
     \left[ C^\star (\bm \eta) \right]^2
\end{equation} 
\begin{equation} \label{igriem2} 
     \Rightarrow \qquad 
     d t^\star = 
     \sqrt{ 
          \left[ {C^\star} (\bm \eta) \right]^{-2} 
          {g^\star}^{ij} (\bm \eta) 
          d \eta_i d \eta_j 
     }.
\end{equation}
which is comparable to the optical metric in Eq. \eqref{conform} 
and $C^\star (\bm \eta) $ is the corresponding conformal 
refractive index factor. 
Furthermore, using Eq. \eqref{igriem2} and the second gradient 
flow equation Eq. \eqref{gradient1}, we can write: 
\begin{equation} \label{igaction2} 
     d \Psi^\star = 
     \frac{\partial \Psi^\star}{\partial \eta_j} d \eta_i = 
     - {g^\star}^{ij} (\bm \eta) 
     \frac{d \eta_j}{d t^\star} 
     d \eta_i = 
     - \sqrt{ 
          \left[ C^\star (\bm \eta) \right]^2 
          {g^\star}^{ij} (\bm \eta) 
          d \eta_i d \eta_j 
     } = 
     - \left[ C^\star (\bm \eta) \right]^2 
     d t^\star.
\end{equation} 
The null metric we can determine from the above optical 
metric Eq. \eqref{igriem2} is: 
\begin{equation} \label{ignull2} 
     d {s^\star}^2 = 
          {g^\star}^{ij} (\bm \eta) d \eta_i d \eta_j - 
          \left[ C^\star (\bm \eta) \right]^2 
          d {t^\star}^2 
      = 0. 
\end{equation} 
If we define a Lagrangian from the metric Eq. \eqref{igriem2} 
according to Eq. \eqref{action}, then we will have: 
\begin{equation} \label{iglag2} 
     L^\star(\bm \eta, \dot{\bm \eta})  := \frac{d t^\star}{d \tau} = 
     \sqrt{ 
          \left[ {C^\star} (\bm \eta) \right]^{-2} 
          {g^\star}^{ij} (\bm \eta) 
          \dot \eta_i \dot \eta_j 
     } = 
     \sqrt{ 
          {G^\star}^{ij} (\bm \eta) 
          \dot \eta_i \dot \eta_j 
     }, 
\end{equation}
where we introduced the conformal Riemannian metric
\begin{equation} \label{Gsij}
 {G^\star}^{ij} (\bm \eta) :=   \left[ {C^\star} (\bm \eta) \right]^{-2} 
          {g^\star}^{ij} (\bm \eta),
\end{equation}
and again, according to Eq. \eqref{potgrad} and Eq. \eqref{gradient1}, 
the canonical momentum ${p\star}^i$ is given by: 
$$
     \theta^i = \frac{\partial \Psi^\star}{\partial \eta_i} 
\quad \Rightarrow \quad 
     \frac{d \eta_i}{d t^\star} = 
     - g^\star_{ij} (\bm \eta) 
     \frac{\partial \Psi^\star}{\partial \eta_j} = 
     - g^\star_{ij} (\bm \eta) \theta^j
$$
\begin{equation} \label{igmom2} 
     {p^\star}^i := 
     \frac{\partial L^\star}{\partial \dot \eta_i} = 
          \left[ {C^\star} (\bm \eta) \right]^{-2} 
          {g^\star}^{ij} (\bm \eta) 
          \frac{d \eta_j}{d t^\star} = 
          - \left[ {C^\star} (\bm \eta) \right]^{-2} 
          \frac{\partial \Psi^\star}{\partial \eta_i} = 
          - \left[ {C^\star} (\bm \eta) \right]^{-2} \theta^i, 
\end{equation} 
and by Maupertuis principle Eq. \eqref{maup}, we can write 
the Maupertuis form of the metric Eq. \eqref{igriem2} as: 
\begin{equation} \label{igmaup2} 
     d t^\star = 
     {p^\star}^i \; d \eta_i = 
     - \left[ {C^\star} (\bm \eta) \right]^{-2} 
     \theta^i \; d \eta_i
\quad \Rightarrow \quad 
     d \Psi^\star = \theta^i \; d \eta_i. 
\end{equation} 
Furthermore, if we apply Eq. \eqref{potgrad} and Eq. \eqref{igmom2} 
to Eq. \eqref{norm2}, then we can formulate a constraint:
$$
     \left[ {C^\star} (\bm \eta) \right]^{-2}
     {g^\star}^{ij} (\bm \eta) 
     \frac{d \eta_i}{d t^\star} 
     \frac{d \eta_j}{d t^\star} = 
     \frac1{\left[ C^\star (\bm \eta) \right]^2} 
     g^\star_{ij} (\bm \eta) 
     \theta^i \theta^j = 1.
$$
\begin{equation} \label{igham2} 
     \Phi^\star (\bm{\eta, \theta}) = 
     \sqrt{
          {G^\star}_{ij} (\bm \eta) {p^\star}^i {p^\star}^j
     } = 
     \sqrt{
          \left[ C^\star (\bm \eta) \right]^2 
          g^\star_{ij} (\bm \eta) {p^\star}^i {p^\star}^j
     } = 
     \sqrt{
          \frac1{\left[ C^\star (\bm \eta) \right]^2} 
          g^\star_{ij} (\bm \eta) \theta^i \theta^j
     } = 1, 
\end{equation} 
\begin{equation} \label{igcons2} 
     \frac{d \Phi^\star}{d t^\star} = 
     \frac{\partial \Phi^\star}{\partial \theta^i} 
     \frac{d \theta^i}{d t^\star} + 
     \frac{\partial \Phi^\star}{\partial \eta_i} 
     \frac{d \eta_i}{d t^\star} = 0
\quad \Rightarrow \quad 
     \frac{\partial \Phi^\star}{\partial \eta_i} 
     \frac{d \eta_i}{d t^\star} = 
     \frac1{\left[ C^\star (\bm \eta) \right]^2} 
     \frac{d \theta^i}{d t^\star}
     \frac{d \eta_i}{d t^\star}, 
\end{equation} 
which is essentially the alternative constraint, that generates 
the constraint equations: 
\begin{equation} \label{igconseq21} 
     \frac{\partial \Phi^\star}{\partial \theta^i} = 
     \frac1{\left[ C^\star (\bm \eta) \right]^2} 
     g^\star_{ij} (\bm \eta) 
     \theta^j = 
     - \frac1{\left[ C^\star (\bm \eta) \right]^2} 
     \frac{d \eta_i}{d t^\star} = 
     - \frac{d \eta_i}{d \lambda^\star}, 
\qquad \text{where } \ 
     d \lambda^\star = 
     \left[ C^\star (\bm \eta) \right]^2 dt^\star, 
\end{equation} 
\begin{equation} \label{igconseq22} 
     \Rightarrow \qquad 
     \frac{d \theta^i}{d \lambda^\star} = 
     \frac1{\left[ C^\star (\bm \eta) \right]^2} 
     \frac{d \theta^i}{d t^\star} = 
     \frac{\partial \Phi^\star}{\partial \eta_i} = 
     \frac12 \frac{\partial \ }{\partial \eta_i} 
     \left( 
          \frac{g^\star_{ij} (\bm \eta)}{\left[ C^\star (\bm \eta) \right]^2}  
     \right) 
     \theta^i \theta^j. 
\end{equation} 
Furthermore, by applying the metric rule Eq. \eqref{metric}, 
if we treat $\bm \theta$ as a function of $\bm \eta$ in 
Eq. \eqref{igcons2} and apply Eq. \eqref{igconseq21} and 
Eq. \eqref{igconseq22}, we can also write: 
$$
     \left( 
          \frac{\partial \Phi^\star}{\partial \theta^j}
          \frac{\partial \theta^j}{\partial \eta_i} + 
          \frac{\partial \Phi^\star}{\partial \eta_i} 
     \right) 
     \frac{d \eta_i}{d t} = 0 
\quad \Rightarrow \quad 
     \frac{\partial \Phi^\star}{\partial \eta_i} = 
     - {g^\star}^{ij} (\bm \eta) 
     \frac{\partial \Phi^\star}{\partial \theta^j} = 
     - \frac1{\left[ C^\star (\bm \eta) \right]^2} 
     \theta^i, 
$$ 
\begin{equation} \label{rule2} 
     \Rightarrow \qquad 
     \frac12 \frac{\partial \ }{\partial \eta_i} 
     \left( 
          \frac{g^\star_{jk} (\bm \eta) }{\left[ C^\star (\bm \eta) \right]^2} 
     \right) 
     \theta^j \theta^k = 
     - \frac1{\left[ C^\star (\bm \eta) \right]^2} 
     \theta^i,
\end{equation}  
\begin{equation} \label{iggradeq2} 
     \therefore \qquad 
     \frac{d \theta^i}{d \lambda^\star} = 
     \frac1{\left[ C^\star (\bm \eta) \right]^2} 
     \frac{d \theta^i}{d t} = 
     \frac{\partial \Phi^\star}{\partial \eta_i} = 
     - \frac1{\left[ C^\star (\bm \eta) \right]^2} 
     \theta^i 
\quad \Rightarrow \quad 
     \boxed{\frac{d \theta^i}{dt^\star} = - \theta^i},
\end{equation}
which is the second equations of Eq. \eqref{gradient3}. 
In this way we have related the first gradient-flow 
to the geodesic flow described by the Lagrangian 
$L^\star({\bm \eta}, \dot{ \bm \eta})$ 
in Eq. \eqref{iglag2} with the conformal Riemannian 
metric ${G^\star}^{ij}(\bm \eta)$ in Eq. \eqref{Gsij}.

\subsection{Canonical Transformation} 
\label{sec:cantransform} 

In classical mechanics, the canonical transformation allows 
us to rewrite the mechanical system by performing a co-ordinate 
transformation between the phase space co-ordinates. 
The most frequently discussed examples involve switching 
the roles of co-ordinate and momenta. 
Here, we will briefly discuss how such canonical transformations 
could possibly allow us to connect the two dual dynamical systems 
studied in IG. 
\bigskip 

\noindent 
Recalling the canonical transformation discussed in Subsec. 
\ref{sec:cmrf}, we can say that the equivalent of Eq. \eqref{canonical} 
here in IG will be: 
$$
     G (\bm{\theta, \eta}) = 
     \eta_i \, \theta^i: \ 
     \left( \theta^i, \eta_i \right) 
     \longrightarrow 
     \left( \eta_i, - \theta^i \right),
$$
\begin{equation} \label{igcanon1} 
\Rightarrow \qquad 
     {{\widehat \theta}}^i = \eta_i,
\quad , \quad 
     {\widehat \eta}_i = - \theta^i.
\end{equation} 
Now, if we apply this transformation rule to the metric rule 
Eq. \eqref{metric} and the first potential gradient formula Eq. \eqref{potgrad}, 
then we shall have: 
\begin{equation} \label{canmet} 
     g^{ij} (\bm \theta) = 
     \frac{\partial \theta^i}{\partial \eta_j} = 
     \frac{\partial \left( - {\widehat \eta}_i \right)}{\partial {{\widehat \theta}}^j \ \ } = 
     - g^\star_{ij} (\bm {\widehat \eta}),
\end{equation}
\begin{equation} \label{canpot1} 
     \eta_i = \frac{\partial \Psi}{\partial \theta^i} 
\quad \longrightarrow \quad 
     {{\widehat \theta}}^i = - \frac{\partial \Psi}{\partial {\widehat \eta}_i}, 
\end{equation} 
where the transformation  Eq. \eqref{canpot1} is similar to Eq. \eqref{canmaup}. 
To fit Eq. \eqref{canpot1} into the form of the second gradient 
in Eq. \eqref{potgrad}, let us first define a second pair of functions 
$\xi ({\widehat \theta})$ and $\xi^\star ({\widehat \eta})$, such that they mirror 
Eq. \eqref{legendre}: 
\begin{equation} \label{newpair} 
     \xi (\bm {\widehat \theta}) = 
     - \Psi^\star (\bm \eta) 
\quad , \quad 
     \xi^\star (\bm {\widehat \eta}) = 
     - \Psi (\bm \theta) 
\quad , \quad 
     \xi (\bm {\widehat \theta}) + 
     \xi^\star (\bm {\widehat \eta}) = 
     {\widehat \eta}_i 
     {{\widehat \theta}}^i,
\end{equation}  
Thus, using Eq. \eqref{newpair} we shall write, applying Eq. \eqref{igaction1}: 
\begin{equation} \label{igcanon2} 
     d t = 
     \left[ C (\bm \theta) \right]^2 d \Psi = 
     - \left[ C (\bm {\widehat \eta}) \right]^2 d \xi^\star = 
     - d t^\star,
\end{equation} 
letting us rewrite Eq. \eqref{canpot1} into: 
\begin{equation} \label{canpot2} 
     {{\widehat \theta}}^i = \frac{\partial \xi^\star}{\partial {\widehat \eta}_i}.
\end{equation} 
Thus, applying Eq. \eqref{igcanon1}, Eq. \eqref{canmet}, and Eq. \eqref{igcanon2} 
to the first gradient-flow equation of Eq. \eqref{gradient1}, we get: 
\begin{equation} \label{canflow} 
     \frac{d {\widehat \eta}_i}{d t^\star} = 
     - g^\star_{ij} (\bm {\widehat \eta}) 
     \frac{\partial \xi^\star}{\partial {\widehat \eta}_j}, 
\end{equation} 
which is the second gradient-flow equation of Eq. \eqref{gradient1}. 
Furthermore, if we apply the canonical transformation Eq. \eqref{igcanon1} 
and Eq. \eqref{igcanon2} to the first gradient-flow equation of Eq. \eqref{gradient3}, 
then we shall get: 
\begin{equation} \label{cangrad} 
     \frac{d \eta_i}{d t} = \eta_i 
\quad \longrightarrow \quad 
     \frac{d {{\widehat \theta}}^i}{d t^\star} = - {{\widehat \theta}}^i, 
\end{equation} 
which is essentially the second gradient-flow equation of Eq. \eqref{gradient3}. 
Furthermore, if we consider the the potential $\Psi (\bm \theta)$ 
as comparable to the action based on comparison between the first gradient 
flow equation of Eq. \eqref{gradient1} and Eq. \eqref{hamjacobi2}, then using 
Eq. \eqref{newpair}, we can write:  
\begin{equation} \label{canpot} 
     \xi (\bm {\widehat \theta}) = 
     \xi (\bm \eta) = 
     {\widehat \eta}_i 
     {{\widehat \theta}}^i - \xi^\star (\bm {\widehat \eta}) = 
     \Psi (\bm \theta) - \eta_i \theta^i = 
     \Psi (\bm \theta) - G (\bm{\theta, \eta}).
\end{equation} 
Taking the differential of Eq. \eqref{canpot}, we can see that 
it fits the form of the canonical transformation described 
by Eq. \eqref{infocan} and Eq. \eqref{canonical}: 
\begin{equation} \label{infotrans} 
     d \xi = 
     d \Psi - 
     d \left( \eta_i \theta^i \right) = 
     d \Psi - d G. 
\end{equation}  
Thus, we have shown that the gradient-flow equations of Eq. \eqref{gradient1} 
are essentially canonical duals to each other.

\numberwithin{equation}{section}

\section{Randers-Finsler metric in IG} 
\label{sec:rfinfo} 

The Randers-Finsler metric can be regarded as a modification 
of the Riemannian metric via the addition of a linear term. 
This suggests that such a metric can be realised in IG via a 
similar deformation wherever suitable. 
There are two types of deformation we shall consider here, 
deformation of the metric, and then of the potential functions. 
\bigskip 

As before in Sec. \ref{sec:iggeodesic}, we shall demonstrate 
the similarity between mechanics in RF geometry and dynamics 
in deformed IG. 
Consider the RF metric Eq. \eqref{rander}: 
$$ 
     ds = 
     \sqrt{
          g_{ij} (\bm x) 
          d x^i d x^j
     } + 
     A_i (\bm x) d x^i. 
$$ 
According to Eq. \eqref{randmaup} we can write from 
the gauge-covariant momentum Eq. \eqref{gcmom} that 
\begin{equation} \label{rfhamjacobi1} 
     \frac{d x^i}{d s} = 
     g^{ij} (\bm x) 
     \left( 
          \frac{\partial s}{\partial x^j} - 
          A_i (\bm x) 
     \right). 
\end{equation} 
Here, if we identify a conformal factor within the metric tensor 
and gauge field in Eq. \eqref{rander} such that 
$g^{ij} (\bm x) = \left[ C (\bm x) \right]^2 h^{ij} (\bm x)$, 
then Eq. \eqref{rfhamjacobi1} shall become: 
\begin{equation} \label{rfhamjacobi2} 
     \frac{d x^i}{d t} = 
     h^{ij} (\bm x) 
     \left( 
          \frac{\partial s}{\partial x^j} - 
          A_i (\bm x) 
     \right), 
\qquad \text{where } \ 
     d t = 
     \left[ C (\bm x) \right]^2 
     d s, 
\end{equation} 
which is comparable to a deformation of the first gradient flow 
equation in Eq. \eqref{gradient1}. 
Based on Eq. \eqref{rfhamjacobi2}, we shall attempt to deform 
the gradient-flow equations Eq. \eqref{gradient1}, and derive the 
corresponding metrics.

\numberwithin{equation}{subsection} 

\subsection{Deforming the metric} 
\label{sec:metricdeform}

One way to create a RF metric is to use a co-ordinate transformation 
akin to the Gullstrand-Painlev\'e transformation \cite{painleve, gullstrand} 
of the Schwarzschild \cite{lemaitre} metric. 
Let us take the first conformal Riemannian metric Eq. \eqref{igriem1} 
derived from the first gradient-flow equation from Eq. \eqref{gradient1}. 
If we modify it to write 
\begin{equation} \label{rfmod11} 
     d t = d \widetilde t - 
     \left[ C (\bm \theta) \right]^{-2} 
     A_i (\bm \theta) d \theta^i, 
\end{equation} 
then we can write the RF metric as: 
\begin{equation} \label{rfmod12} 
     d \widetilde t = 
     \sqrt{ 
          \left[ C (\bm \theta) \right]^{-2} 
          g_{ij} (\bm \theta) 
          d \theta^i d \theta^j 
     } + 
     \left[ C(\bm \theta) \right]^{-2} 
     A_i (\bm \theta) d \theta^i. 
\end{equation} 
Naturally, we can deduce the momentum from this metric 
Eq. \eqref{rfmod12} to be: 
\begin{equation} \label{momrf1} 
     \widetilde p_i = 
     \left[ C (\bm \theta) \right]^{-2} 
     \left( 
          g_{ij} (\bm \theta) 
          \frac{d \theta^j}{d t} + 
          A_i (\bm \theta)
     \right). 
\end{equation} 
Using the first gradient-flow equation of Eq. \eqref{gradient1}, 
based on comparison to Eq. \eqref{igmom1}, we can write from 
Eq. \eqref{momrf1}: 
$$
     \widetilde p_i = 
     \left[ C (\bm \theta) \right]^{-2} 
     \left( 
          \frac{\partial \Psi}{\partial \theta^i} + 
          A_i (\bm \theta)
     \right) = 
     \left[ C (\bm \theta) \right]^{-2} 
     \frac{\partial \widetilde \Psi}{\partial \theta^i}. 
$$
\begin{equation} \label{deform1} 
     \frac{\partial \widetilde \Psi}{\partial \theta^i} = 
     \frac{\partial \Psi}{\partial \theta^i} + 
     A_i (\bm \theta) = 
     \eta_i + 
     A_i (\bm \theta). 
\end{equation} 
Similarly, we can do the same for the second conformal metric 
Eq. \eqref{igriem2} by writing: 
\begin{equation} \label{rfmod21} 
     d t^\star = d \widetilde t^\star - 
     \left[ C^\star (\bm \eta) \right]^{-2} 
     {A^\star}^i (\bm \eta) d \eta_i, 
\end{equation} 
\begin{equation} \label{rfmod22} 
     d \widetilde t^\star = 
     \sqrt{ 
          \left[ C^\star (\bm \eta) \right]^{-2} 
          {g^\star}^{ij} (\bm \eta) 
          d \eta^i d \eta^j 
     } + 
     \left[ C^\star (\bm \eta) \right]^{-2} 
     {A^\star}^i (\bm \eta) d \eta^i, 
\end{equation} 
Thus, giving us the modified potential $\widetilde \Psi^\star$: 
\begin{equation} \label{momrf2} 
     \widetilde {p^\star}^i = 
     \left[ C^\star (\bm \eta) \right]^{-2} 
     \left( 
          {g^\star}^{ij} (\bm \eta) 
          \frac{d \eta_j}{d t^\star} + 
          {A^\star}^i (\bm \eta)
     \right) = 
     \left[ C^\star (\bm \eta) \right]^{-2} 
     \frac{\partial \widetilde \Psi^\star}{\partial \eta_i}. 
\end{equation} 
\begin{equation} \label{deform2} 
     \frac{\partial \widetilde \Psi^\star}{\partial \eta_i} = 
     \frac{\partial \Psi^\star}{\partial \eta_i} + 
     {A^\star}^i (\bm \eta) = 
     \theta^i + 
     {A^\star}^i (\bm \eta). 
\end{equation} 
However, in this case, the original gradient-flow equation from 
Eq. \eqref{gradient1} was not disturbed, along with equations 
Eq. \eqref{potgrad} and Eq. \eqref{metric}. 
This means that the gradient-flow equations Eq. \eqref{gradient3} shall 
remain undisturbed. 
We shall next consider what happens if we deform the gradient 
flow equation Eq. \eqref{gradient1} themselves.

\subsection{Deforming the gradient-flow equation} 
\label{sec:flowdeform}

Since the origin of our formulation are the gradient flow 
equations Eq. \eqref{gradient1}, we shall start by considering 
a deformation of the potential functions. 
\bigskip 

\noindent
If we can choose to deform the gradient-flow equations 
Eq. \eqref{gradient1} as follows: 
\begin{equation} \label{igdeform} 
     \frac{d \theta^i}{d t} = 
     g^{ij} (\bm \theta) 
     \left( 
          \frac{\partial \Psi}{\partial \theta^j} - 
          A_j (\bm \theta) 
     \right) = 
     \chi^i (\bm \theta) 
\quad , \quad  
     \frac{d \eta_i}{d t^\star} = 
     - g^\star_{ij} (\bm \eta) 
     \left( 
          \frac{\partial \Psi^\star}{\partial \eta_j} + 
          {A^\star}^j (\bm \eta) 
     \right) = 
     \chi^\star_i (\bm \eta) . 
\end{equation} 
It automatically follows that upon applying the potential gradient 
Eq. \eqref{potgrad} and the metric rule Eq. \eqref{metric} to the deformed 
gradient-flow equations Eq. \eqref{igdeform}, the gradient-flow equations 
Eq. \eqref{gradient3} are replaced by: 
\begin{equation} \label{graddeform} 
     \boxed{
          \frac{d \eta_i}{d t} = 
          \eta_i - A_i (\bm \theta) 
     \quad , \quad 
          \frac{d \theta^i}{d t^\star} = 
          - \theta^i - {A^\star}^i (\bm \eta).
     }
\end{equation}

\subsubsection*{Deformation of the first gradient-flow equation} 

From the first equation of Eq. \eqref{igdeform}, in the same manner 
that Eq. \eqref{igriem1} was derived in Subsec \ref{sec:inforiem}, 
we can write: 
\begin{equation} \label{igrf11} 
     d t = 
     \sqrt{ 
          \widetilde G_{ij} (\bm \theta) 
          d \theta^i d \theta^j 
     } = 
     \sqrt{ 
          \left[ \widetilde C (\bm \theta) \right]^{-2} 
          g_{ij} (\bm \theta) 
          d \theta^i d \theta^j 
     }, 
\end{equation} 
where $\left[ \widetilde C (\bm \theta) \right]^2 = 
g_{ij} (\bm \theta) \chi^i (\bm \theta) \chi^j (\bm \theta)$. 
Now, in the same manner that Eq. \eqref{igaction1} was derived 
in Subsec \ref{sec:inforiem}, we can write using the first equation 
of Eq. \eqref{igdeform} and Eq. \eqref{igrf11}: 
$$
     d \Psi = 
     \frac{\partial \Psi}{\partial \theta^i} 
     d \theta^i = 
     g_{ij} (\bm \theta) 
     \frac{d \theta^j}{d t} 
     d \theta^i + 
     A_i (\bm \theta) 
     d \theta^i,
$$
\begin{equation} \label{igrfset1} 
\Rightarrow \qquad 
     d \Psi = 
     \left[ \widetilde C (\bm \theta) \right]^2 
     d t + 
     A_i (\bm \theta) 
     d \theta^i = 
     \left[ \widetilde C (\bm \theta) \right]^2 
     d \widetilde t,
\end{equation} 
then the RF metric can be written as: 
\begin{equation} \label{igrf12} 
     d \widetilde t = d t + 
     \left[ \widetilde C (\bm \theta) \right]^{-2} 
     A_i (\bm \theta) 
     d \theta^i. 
\end{equation} 
From this metric Eq. \eqref{igrf12}, using the first equations 
of Eq. \eqref{potgrad} and the Eq. \eqref{igdeform}, we can write 
the gauge covariant momentum $\Pi$ as: 
\begin{equation} \label{rfmom1} 
     \widetilde p_i = 
     \left[ \widetilde C (\bm \theta) \right]^{-2} 
     \left[ 
          g_{ij} (\bm \theta) 
          \frac{d \theta^j}{dt} + 
          A_i (\bm \theta) 
     \right] = 
     \left[ \widetilde C (\bm \theta) \right]^{-2} 
     \eta_i, 
\end{equation} 
\begin{equation} \label{covmom1} 
     \Pi_i = 
     \widetilde p_i - 
     \left[ \widetilde C (\bm \theta) \right]^{-2} 
     A_i (\bm \theta) = 
     \frac1{\left[ C (\bm \theta) \right]^2} 
     \pi_i = 
     \left[ \widetilde C (\bm \theta) \right]^{-2} 
     g_{ij} (\bm \theta) 
     \frac{d \theta^j}{dt}
\end{equation} 
where $\pi_i = \eta_i - A_i (\bm \theta)$. 
The constraint for this RF metric Eq. \eqref{igrf12} is given 
by using Eq. \eqref{covmom1} as: 
\begin{equation} \label{igrfham1} 
     \Phi (\bm{\eta, \theta}) = 
     \sqrt{
          G^{ij} (\bm \theta) 
          \Pi_i \Pi_j
     } = 
     \sqrt{
          \frac1{\left[ \widetilde C (\bm \theta) \right]^2} 
          g^{ij} (\bm \theta) 
          \pi_i \pi_j 
     } = 1, 
\end{equation} 
that generates the following constraint equations: 
\begin{equation} \label{igrfconseq11} 
     \frac{\partial \Phi}{\partial \eta_i} = 
     \frac1{\left[ \widetilde C (\bm \theta) \right]^2} 
     g^{ij} (\bm \theta) 
     \pi_j = 
     \frac{d \theta^i}{d \lambda}, 
\quad \text{where } \ 
     d \lambda := 
     \left[ \widetilde C (\bm \theta) \right]^2 d t
\end{equation} 
\begin{equation} \label{igrfconseq12} 
     \frac{d \eta_i}{d \lambda} = 
     \frac1{\left[ \widetilde C (\bm \theta) \right]^2} 
     \frac{d \eta_i}{d t} = 
     - \frac{\partial \Phi}{\partial \theta^i} = 
     - \frac12 \frac{\partial \ }{\partial \theta^i} 
     \left( 
          \frac{g^{jk} (\bm \theta) }{\left[ \widetilde C (\bm \theta) \right]^2} 
     \right)
     \pi_j \pi_k + 
     \frac{d \theta^j}{d \lambda} 
     \frac{\partial A_j}{\partial \theta^i}. 
\end{equation}

\subsubsection*{Deformation of the second gradient-flow equation} 

In a similar manner, from the second equation of Eq. \eqref{igdeform}, 
we can write: 
\begin{equation} \label{igrf21} 
     d t^\star = 
     \sqrt{ 
          {\widetilde G}^{\star ij} (\bm \eta) 
          d \eta_i d \eta_j 
     } = 
     \sqrt{ 
          \left[ \widetilde C^\star (\bm \eta) \right]^{-2} 
          {g^\star}^{ij} (\bm \eta) 
          d \eta_i d \eta_j 
     }, 
\end{equation} 
where $\left[ \widetilde C^\star (\bm \eta) \right]^2 = 
     {g^\star}^{ij} (\bm \eta) \chi^\star_i \chi^\star_j$. 
As before for Eq. \eqref{igaction2} in Subsec \ref{sec:inforiem}, 
we can write using the second equation of Eq. \eqref{igdeform} 
and Eq. \eqref{igrf21}: 
$$
     d \Psi^\star = 
     \frac{\partial \Psi^\star}{\partial \eta_i} 
     d \eta_i = 
     - \left( 
          {g^\star}^{ij} (\bm \eta) 
          \frac{d \eta_j}{d t^\star} 
          d \eta_i + 
          {A^\star}^i (\bm \eta) 
          d \eta_i
     \right),
$$
\begin{equation} \label{igrfset2} 
\Rightarrow \qquad 
     d \Psi^\star = 
     - \left( 
     \left[ 
          \widetilde C^\star (\bm \eta) \right]^2 
          d t^\star + 
          {A^\star}^i (\bm \eta) 
          d \eta_i 
     \right) = 
     - \left[ \widetilde C^\star (\bm \eta) \right]^2 
     d \widetilde t^\star,
\end{equation} 
then the RF metric can be written as: 
\begin{equation} \label{igrf22} 
     d \widetilde t^\star = d t^\star + 
     \left[ \widetilde C^\star (\bm \eta) \right]^{-2} 
     {A^\star}^i (\bm \eta) 
     d \eta_i. 
\end{equation} 
From this metric Eq. \eqref{igrf22}, using the second equations 
of Eq. \eqref{potgrad} and the Eq. \eqref{igdeform}, we can write 
the gauge covariant momentum $\Pi^\star$ as: 
\begin{equation} \label{rfmom2} 
     \widetilde {p^\star}^i = 
     \left[ \widetilde C^\star (\bm \eta) \right]^{-2} 
     \left[ 
          {g^\star}^{ij} (\bm \eta) 
          \frac{d \eta_j}{d t^\star} + 
          {A^\star}^i (\bm \eta) 
     \right] = 
     - \left[ \widetilde C^\star (\bm \eta) \right]^{-2} 
     \theta^i, 
\end{equation} 
\begin{equation} \label{covmom2} 
     {\Pi^\star}^i = 
     \widetilde {p^\star}^i - 
     \left[ \widetilde C^\star (\bm \eta) \right]^{-2} 
     {A^\star}^i (\bm \eta) = 
     \frac1{\left[ C^\star (\bm \eta) \right]^2} 
     {\pi^\star}^i = 
     \left[ \widetilde C^\star (\bm \eta) \right]^{-2} 
     {g^\star}^{ij} (\bm \eta) 
     \frac{d \eta_j}{d t^\star}.
\end{equation} 
where ${\pi^\star}^i = - \theta^i - {A^\star}^i$. 
The constraint for this RF metric Eq. \eqref{igrf22} is given 
by using Eq. \eqref{covmom2} as: 
\begin{equation} \label{igrfham2} 
     \Phi^\star (\bm{\eta, \theta}) = 
     \sqrt{
          \left[ \widetilde C^\star (\bm \eta) \right]^2 
          g^\star_{ij} (\bm \eta) 
          {\Pi^\star}^i {\Pi^\star}^j
     } = 
     \sqrt{
          \frac1{\left[ \widetilde C^\star (\bm \eta) \right]^2} 
          g^\star_{ij} (\bm \eta) 
          {\pi^\star}^i {\pi^\star}^j 
     } = 1, 
\end{equation} 
that generates the following constraint equations: 
\begin{equation} \label{igrfconseq21} 
     \frac{\partial \Phi^\star}{\partial \theta^i} = 
     - \frac1{\left[ \widetilde C^\star (\bm \eta) \right]^2} 
     g^\star_{ij} (\bm \eta) 
     {\pi^\star}^j = 
     - \frac{d \eta_i}{d \lambda^\star}, 
\quad \text{where } \ 
     d \lambda^\star := 
     \left[ \widetilde C^\star (\bm \eta) \right]^2 d t^\star,
\end{equation} 
\begin{equation} \label{igrfconseq22} 
          \frac{d \theta^i}{d \lambda^\star} = 
          \frac1{\left[ \widetilde C^\star (\bm \eta) \right]^2} 
          \frac{d \theta^i}{d t^\star} = 
          \frac{\partial \Phi^\star}{\partial \eta_i} = 
          \frac12 \frac{\partial \ }{\partial \eta_i} 
          \left( 
               \frac{g^\star_{jk} (\bm \eta) }{\left[ \widetilde C^\star (\bm \eta) \right]^2} 
          \right)
          {\pi^\star}^j {\pi^\star}^k - 
          \frac{d \eta_j}{d \lambda^\star} 
          \frac{\partial {A^\star}^j}{\partial \eta_i}. 
\end{equation} 
With this, we conclude our discussion on Randers-Finsler 
metrics in IG. We will finally proceed to the last stage which 
is the application and testing of our theories on one example 
known as the Gaussian model.

\numberwithin{equation}{section} 

\section{Application to the Gaussian Model}
\label{sec:gauss}

Having established our theory, we will now verify it by applying our theory 
to the Gaussian or Normal distribution model $N (\mu, \sigma)$ as a well 
known example in IG. 
\begin{equation} \label{normal} 
     P_G (x; \mu, \sigma^2) = 
     \frac1{\sqrt{2 \pi \sigma^2}} 
     \text{exp } 
     \left[ 
          - \frac{\left( x - \mu \right)^2}{2 \sigma^2} 
     \right], 
\end{equation} 
where $\mu$ and $\sigma^2$ are the mean and dispersion respectively. 
If we define the associated natural co-ordinate system as a modification 
of the Gaussian model described in \cite{wsm2}: 
\begin{equation} \label{gaussvar} 
     \begin{split} 
          \eta_1 = \mu 
     \qquad &, \qquad 
          \eta_2 = \mu^2 + \sigma^2 ,
     \\ 
          \theta^1 = \frac{\mu}{\sigma^2} 
     \qquad &, \qquad 
          \theta^2 = - \frac1{2 \sigma^2} .
     \end{split}
\end{equation} 
According to the metric rules Eq. \eqref{metric}, we have the metric 
${g^\star}^{ij} (\bm \eta)$ and the inverse metric $g^\star_{ij} (\bm \eta)$ 
given from Eq. \eqref{gaussvar} by: 
\begin{equation} \label{gaussmet} 
     {g^\star}^{ij} (\bm \eta) = 
     \frac{\partial \theta^i}{\partial \eta_j} = 
     \frac{\partial \theta^j}{\partial \eta_i} = 
     \frac1{\left( \eta_2 - {(\eta_1)}^2 \right)^2} 
     \left( \begin{array}{cc} 
          \eta_2 + \left( \eta_1 \right)^2 & - \eta_1 
     \\ 
          - \eta_1 & \frac12 
     \end{array} \right) = 
     \frac1{\sigma^4} 
     \left( \begin{array}{cc} 
          2 \mu^2 + \sigma^2 & - \mu 
     \\ 
          - \mu & \frac12 
     \end{array} \right),
\end{equation} 
\begin{equation} \label{gaussinvmet} 
     g^\star_{ij} (\bm \eta) = 
     \frac{\partial \eta_i}{\partial \theta^j} = 
     \frac{\partial \eta_j}{\partial \theta^i} = 
     \left( 
          \eta_2 - {(\eta_1)}^2
     \right) 
     \left( \begin{array}{cc} 
          1 & 2 \eta_1 
     \\ 
          2 \eta_1 & 2 \left( \eta_2 + {(\eta_1)}^2 \right) 
     \end{array} \right) = 
     \sigma^2 
     \left( \begin{array}{cc} 
          1 & 2 \mu 
     \\ 
          2 \mu & 2 \left( \sigma^2 + 2 \mu^2 \right)
     \end{array} \right).
\end{equation} 
and the derivative of the inverse metric Eq. \eqref{gaussinvmet} 
are given by: 
\begin{equation} \label{metgrads} 
     \frac{\partial g_{ij}}{\partial \eta_1} = 
     \left( 
          \begin{array}{cc} 
               - 2 \mu  &  2 \sigma^2 - 4 \mu^2 \\ 
               2 \sigma^2 - 4 \mu^2  &  - 8 \mu^3
          \end{array} 
     \right) 
\quad , \quad  
     \frac{\partial g_{ij}}{\partial \eta_2} = 
     \left( 
          \begin{array}{cc} 
               1  &  2 \mu \\ 
               2 \mu  &  - 4 \left( \sigma^2 + \mu^2 \right) 
          \end{array} 
     \right), 
\end{equation} 
These formulas will be necessary for both 
the Riemannian metric and RF deformation 
we will study here.

\numberwithin{equation}{subsection}

\subsection{Riemannian metric for Gaussian model} 
\label{sec:riemgauss}

When formulating a Riemannian metric in IG for the Gaussian model, 
the correspondence to phase space co-ordinates is that for the second 
gradient-flow equation of Eq. \eqref{gradient1} in subsection \ref{sec:gradflow} 
and given by: 
$$
\left\{ 
     \begin{split} 
          \eta-\text{co-ordinates:} 
          &\qquad 
          x^i  
          \quad \Leftrightarrow \quad \ \ 
          \eta_i, 
     \\ 
          \theta-\text{co-ordinates:} 
          &\qquad 
          p_i  
          \quad \Leftrightarrow \quad 
          - \left[ {C^\star}^{-1} (\theta) \right]^2 \theta^i. 
     \end{split} \right. 
$$
Knowing $\bm \eta$ and $\bm \theta$, we must immediately 
determine the metric $g^\star_{ij} (\bm \eta)$ and inverse 
${g^\star}^{ij} (\bm \eta)$, and the conformal refraction factor 
$C^\star (\bm \eta)$. 
Doing so will allow us to determine the associated Riemannian 
metric. 
\bigskip 

\noindent 
The conformal refraction factor $C^\star (\bm \eta)$ is given by: 
$$
     \left[ C^\star (\bm \eta) \right]^2 = 
     g^\star_{ij} (\bm \eta) 
     \frac{\partial \Psi^\star}{\partial \eta_i} 
     \frac{\partial \Psi^\star}{\partial \eta_j} = 
     g^\star_{ij} (\bm \eta) 
     \theta^i \theta^j = 
     \frac12,
$$
\begin{equation} \label{gaussfactor} 
     C^\star (\bm \eta) = \frac1{\sqrt{2}} 
\quad \Rightarrow \quad 
     \left[ C^\star (\bm \eta) \right]^{-1} = \sqrt{2}.
\end{equation} 
Using Eq. \eqref{gaussmet} and Eq. \eqref{gaussfactor}, we can write the 
Gaussian Information metric according to Eq. \eqref{igriem2} as: 
\begin{equation} \label{gaussriem} 
     d t^\star = 
     \sqrt{
          4 \left[ 
               \left( 
                    (\theta^1)^2 - \theta^2 
               \right) 
               \left( d \eta_1 \right)^2 + 
               2 \theta^1 \theta^2 \; 
               d \eta_1 d \eta_2 + 
               (\theta^2)^2 
               \left( d \eta_2 \right)^2 
          \right] 
     } = 
     \sqrt{
          \frac{2}{\sigma^2} 
          \left( 
               d \mu^2 + 2 d \sigma^2 
          \right) 
     },
\end{equation} 
and the Gaussian model according to Eq. \eqref{ignull2} derives 
from the null metric: 
\begin{equation} \label{orgmet1} 
     d s^\star = 
     \sqrt{
          2 \left[ 
               \left( 
                    (\theta^1)^2 - \theta^2 
               \right) 
               \left( d \eta_1 \right)^2 + 
               2 \theta^1 \theta^2 \; 
               d \eta_1 d \eta_2 + 
               (\theta^2)^2 
               \left( d \eta_2 \right)^2 
          \right] - 
          \frac12 {d t^\star}^2
     } = 0,
\end{equation} 
which, upon applying Eq. \eqref{gaussvar}, is also written as: 
\begin{equation} \label{orgmet2} 
     \boxed{
          d s^\star = 
          \sqrt{
               \frac1{\sigma^2} 
               \left( 
                    d \mu^2 + 2 d \sigma^2 
               \right) - 
               \frac12 {d t^\star}^2
          } = 0.
     }
\end{equation} 
The potential function $\Psi^\star (\bm \eta)$ is given by: 
$$
     d \Psi^\star = 
     \theta^i d \eta_i = 
     \frac{\mu}{\sigma^2} 
     d \mu - 
     \frac1{\sigma^2} \; 
     \left( 
          \mu \; d \mu + 
          \sigma \; d \sigma 
     \right) = 
     - \frac{d \sigma}{\sigma} = 
     - \frac12 d \left( \ln \sigma^2 \right) 
$$
\begin{equation} \label{gausspot} 
     \Psi^\star (\bm \eta) = 
     - \frac12 \ln \sigma^2 = 
     - \frac12 \ln 
     \left(
          \eta_2 - (\eta_1)^2
     \right). 
\end{equation} 
According to Eq. \eqref{igaction2} we shall have: 
\begin{equation} \label{gaussmaup} 
     d t^\star = 
     - \left[ 
          C^\star (\bm \theta) 
     \right]^{-2}
     \theta^i d \eta_i = 
     - 2 \; d \Psi^\star = 
     d \left( \ln \sigma^2 \right). 
\end{equation} 
Thus, we can also formulate a constraint according to Eq. \eqref{igham2}, 
should be: 
\begin{equation} \label{gausscnstr} 
     \Phi^\star (\bm{\eta, \theta}) = 
     \sqrt{
          2 g^\star_{ij} (\bm \eta) \theta^i \theta^j
     } = 1.
\end{equation} 
Using Eq. \eqref{gausscnstr}, we can see that from the first constraint 
equation Eq. \eqref{igconseq21}: 
\[ 
     \begin{split} 
          \frac{d \eta_1}{d \lambda^\star} &= 
          - \frac{\partial \Phi^\star}{\partial \theta^1} = 
          - 2 g_{1j} (\bm \eta) \theta^j = 
          - 2 \sigma^2 
          \frac{\mu}{\sigma^2} + 
          4 \mu \sigma^2 \frac1{2 \sigma^2} = 0
     \quad \Rightarrow \quad 
          \frac{d \mu}{d \lambda^\star} = 0,
     \\ 
          \frac{d \eta_2}{d \lambda^\star} &= 
          - \frac{\partial \Phi^\star}{\partial \theta^2} = 
          - 2 g_{2j} (\bm \eta) \theta^j = 
          - 4 \mu \sigma^2 
          \frac{\mu}{\sigma^2} + 
          4 \sigma^2 
          \left( 
               \sigma^2 + 2 \mu^2 
          \right) 
          \frac1{2 \sigma^2} = 2 \sigma^2 
     \quad \Rightarrow \quad 
          \frac{d \sigma}{d \lambda^\star} = \sigma ,
     \end{split} 
\]
\begin{equation} \label{gvel} 
     \therefore \qquad 
     d \lambda^\star = \frac12 d t^\star 
\quad \Rightarrow \quad  
     \boxed{
          \frac{d \mu}{d t^\star} = 0 
     \quad , \quad  
          \frac{d \sigma}{d t^\star} =  \frac{\sigma}2 .
     }
\end{equation} 
We can therefore verify the 2nd gradient-flow equation of Eq. \eqref{gradient3} 
using Eq. \eqref{gaussvar} and Eq. \eqref{gaussmaup}:  
\begin{equation} \label{set1} 
     \begin{split} 
          \frac{d \theta^1}{d t^\star} &= 
          - 2 \frac{\mu}{\sigma^3} 
          \frac{d \sigma}{d t^\star} = 
          - \frac{\mu}{\sigma^2} = 
          - \theta^1 ,
     \\ 
          \frac{d \theta^2}{d t^\star} &= 
          \frac1{\sigma^3} 
          \frac{d \sigma}{d t^\star} = 
          \frac1{2 \sigma^2} = 
          - \theta^2. 
     \end{split} 
\end{equation} 
Furthermore, we can see that the second gradient-flow equation 
of Eq. \eqref{gradient3} is satisfied by the second constraint equation 
Eq. \eqref{igconseq22} by taking the derivative of the inverse metric 
wrt $\eta^i$ Eq. \eqref{metgrads} to verify Eq. \eqref{rule2}: 
\begin{equation} \label{set2} 
     \begin{split} 
          \frac{d \theta^1}{d \lambda^\star} &= 
          \frac{\partial g_{ab}}{\partial \eta_1} 
          \theta^a \theta^b = 
          - 2 \frac{\mu}{\sigma^2} = 
          - 2 \theta^1 ,
     \\ 
          \frac{d \theta^2}{d \lambda^\star} &= 
          \frac{\partial g_{ab}}{\partial \eta_2} 
          \theta^a \theta^b = 
          \frac1{\sigma^2} = 
          - 2 \theta^2 .
     \end{split} 
\end{equation} 
From both, Eq. \eqref{set1} and Eq. \eqref{set2}, we see that if according 
to Eq. \eqref{igconseq21} we write $d \lambda^\star = \frac12 d t^\star$, 
then we can write: 
\begin{equation} \label{gaussgrad} 
     d \lambda^\star = \frac12 d t^\star 
\quad \Rightarrow \quad 
     \boxed{ 
          \frac{d \theta^1}{d t^\star} = - \theta^1 
     \quad , \quad 
          \frac{d \theta^2}{d t^\star} = - \theta^2. 
     }
\end{equation} 
which are the second gradient-flow equation of Eq. \eqref{gradient3}. 
Thus, our formulation is verified to be completely consistent throughout.

\subsection{RF metric for the Gaussian model} 
\label{sec:rfgauss} 

Here, we shall modify the Gaussian metric Eq. \eqref{orgmet2} into a RF metric 
using the process shown in Subsec \ref{sec:flowdeform}. 
Suppose the gradient-flow equation for the Gaussian system 
is deformed by a gauge field $\bm A^\star (\bm \eta)$ given by: 
$$
     {A^\star}^i (\bm \eta) d \eta_i = 
     \frac1{\sigma} d \mu = 
     \frac1{\sqrt{\eta_2 - \left( \eta_1 \right)^2}} 
     d \eta_1 ,
$$
\begin{equation} \label{pot} 
     {A^\star}^1 (\bm \eta) = 
     \frac1{\sqrt{\eta_2 - \left( \eta_1 \right)^2}} = 
     \frac1{\sigma} 
\quad , \quad 
     {A^\star}^2 (\bm \eta) = 0.
\end{equation} 
In this case, the conformal refraction factor $\widetilde C^\star (\bm \eta)$ 
is given by: 
\smallskip 
$$ 
     \left[ \widetilde C^\star (\bm \eta) \right]^2 = 
     {g^\star}^{ij} (\bm \eta) 
     \chi_i \chi_j = 
     g^\star_{ij} (\bm \eta) 
     \left( \theta^i + {A^\star}^i (\bm \eta) \right) 
     \left( \theta^j + {A^\star}^j (\bm \eta) \right) = 
     \frac32,
$$
\begin{equation} \label{rfgaussfactor} 
     \widetilde C^\star (\bm \eta) = \sqrt{\frac32} 
\quad \Rightarrow \quad 
     \left[ \widetilde C^\star (\bm \eta) \right]^{-1} = \sqrt{\frac23}.
\end{equation} 
Thus, the RF metric according to Eq. \eqref{igrf22} can be given by: 
\begin{equation} \label{gaussrfmet} 
     d \widetilde t^\star = 
     \sqrt{
          \frac{2}{3 \sigma^2} 
          \left( 
               d \mu^2 + 2 d \sigma^2 
          \right) 
     } + 
     \frac2{3 \sigma} d \mu.
\end{equation} 
The deformed gradient-flow equations according to the second 
equation of Eq. \eqref{graddeform} are given by:
\begin{equation} \label{gaussgraddeform} 
     \frac{d \theta^1}{d \widetilde t^\star} = 
     - \theta^1 - 
     \frac1{\sqrt{\eta_2 - \left( \eta_1 \right)^2}} 
\quad , \quad 
     \frac{d \theta^2}{d \widetilde t^\star} = 
     - \theta^2. 
\end{equation} 
This can be verified by defining the new constraint according 
to Eq. \eqref{igrfham2}: 
\begin{equation} \label{gaussrfcon} 
     \Phi^\star (\bm{\eta, \theta}) = 
     \sqrt{
          \frac23 
          g^\star_{ij} (\bm \eta) 
          \left( \theta^i + {A^\star}^i (\bm \eta) \right) 
          \left( \theta^j + {A^\star}^j (\bm \eta) \right) 
     } = 1,
\end{equation} 
and the first equation of motion after straight-forward calculations 
using Eq. \eqref{gaussvar} and Eq. \eqref{gaussinvmet} similar to what 
we have done previously in Subsec \ref{sec:riemgauss} according 
to the first constraint equation Eq. \eqref{igconseq21} are: 
\[ 
     \begin{split} 
          \frac{d \eta_1}{d \lambda^\star} &= 
          - \frac{\partial \Phi^\star}{\partial \theta^1} = 
          - \frac23 g^\star_{1j} (\bm \eta) 
          \left( \theta^j + {A^\star}^j \right) = 
          - \frac23 \sigma 
     \quad \Rightarrow \quad 
          \frac{d \mu}{d \lambda^\star} = - \frac23 \sigma, 
     \\ 
          \frac{d \eta_2}{d \lambda^\star} &= 
          - \frac{\partial \Phi^\star}{\partial \theta^2} = 
          - \frac23 g^\star_{2j} (\bm \eta) 
          \left( \theta^j + {A^\star}^j \right) = 
          \frac23 \sigma^2 - \frac43 \mu \sigma 
     \quad \Rightarrow \quad 
          \frac{d \sigma}{d \lambda^\star} = \frac13 \sigma, 
     \end{split}
\]
\begin{equation} \label{rfgausseq} 
     d t^\star = \frac23 d \lambda^\star 
\quad \Rightarrow \quad 
     \boxed{
          \frac{d \mu}{d t^\star} = - \sigma 
     \quad , \quad 
          \frac{d \sigma}{d t^\star} = \frac12 \sigma .
     }
\end{equation} 
which is different from Eq. \eqref{gvel} where $\mu$ was a constant 
of motion. 
Thus, we can say that according to Eq. \eqref{rfgausseq}, we have
confirmed the deformation of Eq. \eqref{gaussgrad} as follows: 
\begin{equation} \label{gaussdeform1}
     \begin{split} 
          \frac{d \theta^1}{d t^\star} &= 
          \frac{d}{d t^\star} 
          \left( 
               \frac{\mu}{\sigma^2} 
          \right) = 
          \frac{1}{\sigma^2} \frac{d \mu}{d t^\star} - 
          \frac{2 \mu}{\sigma^3} \frac{d \sigma}{d t^\star} =
          - \frac{1}{\sigma} - \frac{\mu}{\sigma^2} = 
          - {A^\star}^1 (\bm \eta) - \theta^1, 
     \\
          \frac{d \theta^2}{d t^\star} &= 
          \frac{d}{d t^\star} 
          \left( 
               - \frac{1}{2 \sigma^2} 
          \right) = 
          \frac{1}{\sigma^3} 
          \frac{d \sigma}{d t^\star} = 
          \frac{1}{2 \sigma^2} = 
          - \theta^2. 
     \end{split}
\end{equation} 
Finally, to verify the deformed gradient-flow equations 
via constraint through Eq. \eqref{igrfconseq22}, we write: 
\begin{equation} \label{gaussdeform2} 
     \begin{split} 
          \frac{d \theta^1}{d \lambda^\star} &= 
          \frac13 
          \frac{\partial g^\star_{ab}}{\partial \eta_1} 
          \left( \theta^a + {A^\star}^a (\bm \eta) \right) 
          \left( \theta^b + {A^\star}^b (\bm \eta) \right) - 
          \frac{d \eta_a}{d \lambda^\star} 
          \frac{\partial {A^\star}^a}{\partial \eta_1} \\ 
          &= - \frac23 \frac{\mu}{\sigma^2} + \frac23 \frac1{\sigma} = 
          - \frac23 \left( \theta^1 + {A^\star}^1 (\bm \eta) \right) , 
     \\ 
          \frac{d \theta^2}{d \lambda^\star} &= 
          \frac13 
          \frac{\partial g^\star_{ab}}{\partial \eta_2} 
          \left( \theta^a + {A^\star}^a (\bm \eta) \right) 
          \left( \theta^b + {A^\star}^b (\bm \eta) \right) - 
          \frac{d \eta_a}{d \lambda^\star} 
          \frac{\partial {A^\star}^a}{\partial \eta_2} \\ 
          &= \frac1{3 \sigma^2} = 
          - \frac23 \theta^2 
     \end{split} 
\end{equation} 
From Eq. \eqref{gaussdeform2}, we see that if we write 
$d \lambda^\star = \frac32 d t^\star$, then we will have: 
\begin{equation} \label{gaussrfgrad} 
     d \lambda^\star = \frac32 d t^\star 
\quad \Rightarrow \quad 
     \boxed{ 
          \frac{d \theta^1}{d t^\star} = - \left( \theta^1 + {A^\star}^1 (\bm \eta) \right)  
     \quad , \quad 
          \frac{d \theta^2}{d t^\star} = - \theta^2,
     }
\end{equation} 
which are the second gradient-flow equation of Eq. \eqref{graddeform}. 
Thus, our formulation is verified to be completely consistent.

\numberwithin{equation}{section} 

\section{Analysis of Black Hole Thermodynamic Geometry}
\label{sec:thermogeom} 

Ruppeiner \cite{ruppeiner} tried to describe thermodynamics 
via geometry in the context of thermodynamic fluctutation theory. 
He argued that such Riemannian geometry had information about 
the underlying statistical mechanical system. 
Although it is suspected that a connection is inevitable once quantum 
theory is involved, there is still much to explore on the classical level. 

Hawking demonstrated how to study black holes as thermodynamic 
systems which involved no statistical mechanics. 
While such fluctuations that served as the basis for Ruppeiner's theory 
may not be associated with black holes, some important information 
can be extracted via Ruppeiner geometry. 
The blackhole of the Reissner-Nordstr\"om blackhole is flat in any 
dimensions. 

Aman, Bengtsson, and Pidokrajt studied flat metrics for the IG 
description of black hole thermodynamics \cite{abp1, ap, abp2}. 
Here, we will develop upon their work by assuming dynamically 
evolving models of their results and analysing them by applying 
our theories. 
\bigskip 

\noindent 
The potential $\Psi$ in thermodynamics is either the the Energy/Mass 
function which leads to Weinhold geometry, 
\begin{equation} \label{weinhold} 
     \psi_W = M (S, Q) 
\quad , \quad 
     g^W_{ij} (\bm x) = 
     \frac{\partial^2 M}{\partial x^i \partial x^j}
\end{equation} 
or the sign-reversed entropy which leads to Ruppeiner geometry
\begin{equation} \label{ruppeiner} 
     \psi_R = - S (M, Q)
\quad , \quad 
     g^R_{ij} (\bm x) = 
     - \frac{\partial^2 S}{\partial x^i \partial x^j}, 
\end{equation} 
where the Weinhold metric $g^W_{ij}$ is a precursor to Ruppeiner 
metric $g^R_{ij}$. 
The connection between Ruppeiner and Weinhold geometries 
can be derived from the First Law of Black Hole thermodynamics, 
written as: 
\begin{equation} \label{firstlaw} 
     d M = T d S + \text{``work terms"}
\end{equation} 
Upon Taylor expansion about $x$, by matching the 2nd order terms 
on both sides of Eq. \eqref{firstlaw} and applying Eq. \eqref{weinhold} and 
Eq. \eqref{ruppeiner} we can write: 
\begin{equation} \label{connection} 
     d s^2 = 
     g^R_{ij} (\bm x) 
     d x^i d x^j = 
     \frac1T 
     g^W_{ij} (\bm x)
      d x^i d x^j
\end{equation} 
In Black Hole Thermodynamics, there is a preferred set 
of variables, i.e., extensive quantities, additive conserved 
charges, and probability distribution. 
Ruppeiner geometry was constructed in the context 
of thermodynamic fluctuation theory. 
Although such fluctuations may not be associated with 
black holes, some essential information can be extracted 
via Ruppeiner geometry. 
\bigskip 

\noindent 
One possible potential we can devise to generate flat 
information metrics is given by \cite{abp2}: 
\begin{equation} \label{flatpot} 
     \psi (x, y) = x^a f (\sigma) 
\qquad , \quad \text{where } 
     \sigma = x^b y, 
\end{equation} 
which generates flat (i.e. vanishing Ricci scalar) metrics that 
are diagonal depending on the choice of co-ordinates and settings 
of parameters. 
From this potential, identifying the variables as 
$(\eta_1, \eta_2) = (x, y)$, we can write the information metric: 
\begin{equation} \label{bhinfomet} 
\begin{split} 
     d s^2 = 
     x^{a-2} 
     & \left[ 
          \left( 
               a (a - 1) f - 
               b (2a - 1) \sigma f' + 
               b^2 \sigma^2 f'' 
          \right)
          d x^2 \right. + \\ 
          & \qquad \qquad 
          \left. 2 \left( 
               (a + b) f' + 
               b \sigma f'' 
          \right) x^{b + 1} 
          d x \; d y + 
          x^{2 (b + 1)} f''
          d y^2 
     \right]. 
\end{split} 
\end{equation} 
Upon setting $b = - a$ and using the variables $(x, \sigma)$ 
we get the flat Weinhold metric \cite{abp2}: 
\begin{equation} \label{flatmetric1} 
     d s^2 = 
     x^{a-2} 
     \left[ 
          a (a - 1) 
          \left( 
               f - \sigma f' 
          \right)
          d x^2 + 
          x^2 f''
          d \sigma^2 
     \right], 
\end{equation} 
while on the other hand, setting $b = - 1$ and using the variables 
$(\psi, \sigma)$ gives us the flat Ruppeiner metric \cite{abp2}: 
\begin{equation} \label{flatmetric2} 
     d s^2 = 
     \frac{a - 1}{a} 
     \frac{d \psi^2}{\psi^2} + 
     \psi 
     \left( 
          \frac{f''}{f} - 
          \frac{a - 1}{a} 
          \left( \frac{f'}{f} \right)^2 
     \right) 
     d \sigma^2. 
\end{equation} 
We can further write the dual variables as: 
\begin{equation} \label{dualcoor} 
     \theta^x = 
     \frac{\partial \psi}{\partial x} = 
     x^{a - 1} 
     \left( 
          a f (\sigma) + 
          b \sigma f' (\sigma) 
     \right) 
\quad , \quad 
     \theta^y = 
     \frac{\partial \psi}{\partial y} = 
     x^{a + b} f' (\sigma). 
\end{equation} 
Under the circumstances that a different set of variables 
is used, a transformation helps determine the new set 
of dual co-ordinates $\{ \widetilde{\theta}^i, \widetilde{\eta}_i \}$. 
From the gradient-flow equation Eq. \eqref{gradient1}, 
we can write: 
$$
     {g^{ij}}^\star (\bm \eta) 
     \frac{d \eta_i}{d t} 
     \frac{d \eta_j}{d t} = 
     - \frac{\partial \psi}{\partial \eta_i} 
     \frac{d \eta_i}{d t} = 
     - \theta^i 
     \frac{d \eta_i}{d t} 
\quad = \quad 
     \widetilde g^{\star ij} (\bm{\widetilde{\theta}}) 
     \frac{d \widetilde \eta_i}{d t} 
     \frac{d \widetilde \eta_j}{d t} = 
     - \widetilde \theta_a 
     \frac{d \widetilde \eta_a}{d t}, 
$$ 
\begin{equation} \label{transform} 
\Rightarrow \qquad 
     \widetilde \theta^a = 
     \frac{\partial \eta_i}{\partial \widetilde \eta_a} 
     \theta^i.
\end{equation} 
Now, we will apply our theories to the analysis 
of the Kerr and the Reissner-Nordstr\"om black holes.

\numberwithin{equation}{subsection}

\subsection{Kerr Black Hole} 

The Kerr black hole are of interest because they are believed 
to exist as physical objects. 
The fundamental relation for the Kerr Black Hole in $D = 4$ 
dimensions \cite{abp1} is given by: 
\begin{equation} \label{kerrmass} 
     M = 
     \frac{D - 2}{4} 
     S^{\frac{D - 3}{D - 2}} 
     \left( 
          1 + \frac{4 J^2}{S^2} 
     \right)^{\frac{1}{D - 2}} = 
     \frac12 
     \sqrt{ 
          S + \frac{4 J^2}{S} 
     } 
\end{equation} 
where $M$ is the mass and $J$ is the spin of the black hole, 
then from Eq. \eqref{kerrmass} we can determine the entropy $S$: 
\begin{equation} \label{kerrentropy} 
     S (M, J) = 
     2 M^2 
     \left( 
          1 \pm 
          \sqrt{ 
               1 - \sigma^2 
          }
     \right), 
\qquad \text{where } \ 
     \sigma = \frac{J}{M^2}
\end{equation} 
If we define the variables accordingly $(x, y) = (M, J)$, then we will have the map: 
\begin{equation} \label{varmap2} 
     \begin{split} 
          \eta_1 &= 
          M 
     \\ 
          \eta_2 &= 
          J = 
          M^2 \sigma, 
     \end{split} 
\quad \Rightarrow \quad 
     \frac{\partial \eta_i}{\partial \widetilde \eta_a} = 
     \left( 
          \begin{array}{cc}
               1 
               & 
               0 
          \\ 
               2 M \sigma 
               & 
               M^2 
          \end{array} 
     \right) 
\end{equation} 
Thus, according to Eq. \eqref{flatmetric1}, we will have the 
following flat Weinhold metric: 
\begin{equation} \label{kerrwein} 
     d s^2 = 
     4 
     \left( 
          1 \pm \frac{1}{\sqrt{1 - \sigma^2}} 
     \right)
     d M^2 \mp 
     M^2 \frac{2}{\left( 1 - \sigma^2 \right)^{\frac32}}
     d \sigma^2 
\end{equation} 
Using the mass Eq. \eqref{kerrmass} as the potential, 
we can write according to Eq. \eqref{gradsol} and Eq. \eqref{dualcoor}: 
\begin{equation} \label{kerrdualcoor1} 
     \begin{split} 
          \theta^M &= 
          \frac{\partial S}{\partial M} = 
          4 M 
          \left( 
               1 \pm 
               \frac{1}{\sqrt{1 - \sigma^2}} 
          \right) = 
          A \text{e}^{- t} 
     \\ 
          \theta^J &= 
          \frac{\partial S}{\partial J} = 
          \mp \frac{2 \sigma}{\sqrt{1 - \sigma^2}} = 
          B \text{e}^{- t}. 
     \end{split}      
\end{equation} 
where $A$ and $B$ are constants. 
From Eq. \eqref{rndualcoor2}, we can say that: 
\begin{equation} \label{sols2} 
     \sigma = 
     1 + 
     \frac{1}{1 + \left( \frac{B}{2} \right)^2 \text{e}^{- 2t}} 
\quad , \quad 
     M = 
     \frac{A \text{e}^{- 2t}}{4 \left( \sqrt{1 + \left( \frac{B}{2} \right)^2 \text{e}^{- 2t}} \right)}. 
\end{equation} 
Finally, using Eq. \eqref{varmap2} and Eq. \eqref{transform}, 
we can write: 
\begin{equation} \label{kerrdualcoor2} 
\begin{split} 
     \left( 
          \begin{array}{c}
               \widetilde \theta^M 
          \\ 
               \widetilde \theta^\sigma
          \end{array} 
     \right) = 
     \left( 
          \frac{\partial \eta_i}{\partial \widetilde \eta_a} 
     \right)^t 
     \left( 
          \begin{array}{c}
               \theta^M 
          \\ 
               \theta^J
          \end{array} 
     \right) &= \left( 
          \begin{array}{cc}
               1 
               & 
               2 M \sigma 
          \\ 
               0 
               & 
               M^2 
          \end{array} 
     \right) 
     \left( 
          \begin{array}{c}
               4 M 
          \left( 
               1 \pm 
               \frac{1}{\sqrt{1 - \sigma^2}} 
          \right) 
          \\ 
               \mp \frac{2 \sigma}{\sqrt{1 - \sigma^2}} 
          \end{array} 
     \right) \\ 
     &= 
     \left( 
          \begin{array}{c}
               4M \left( 1 \pm \sqrt{1 - \sigma^2} \right) 
          \\ 
               \mp M^2 
               \frac{2 \sigma}{\sqrt{1 - \sigma^2}}
          \end{array} 
     \right)    
\end{split}      
\end{equation}

\subsection{Reissner-Nordstr\"om Black Hole} 

The Ruppeiner geometry of Reissner-Nordstr\"om (RN) black 
hole will be flat in any dimension. 
The potential for the information metric of the RN black hole 
in 4 dimensions will be the Entropy \cite{abp2} given by: 
\begin{equation} \label{rnentropy} 
     S (M, Q) = 
     M^2 
     \left(
          1 + 
          \sqrt{
               1 - \frac{Q^2}{M^2}
          } 
     \right)^2, 
\end{equation} 
with the thermodynamic variables identified as 
$(x, y) = (M, Q)$. 
If we identify the potential as the mass $M$, 
\begin{equation} \label{rnmass} 
     M (S, u) = 
     \frac{\sqrt{S}}{2} 
     \left( 
          1 + u^2 
     \right), 
\qquad \text{where } \ 
     u = \frac{Q}{\sqrt{S}} \in \left(-1, 1 \right)
\end{equation} 
and we instead use the variables 
$(x, y) = (S, u)$ with $b = - \frac12$, 
then upon applying to Eq. \eqref{flatmetric2} first leads 
to the flat Weinhold metric according to Eq. \eqref{flatmetric1} 
\cite{abp1}: 
\begin{equation} \label{rnwein} 
     d s^2_W = 
     \frac1{8 S^{\frac{3}{2}}} 
     \left[ 
          - \left( 1 - u^2 \right) 
          d S^2 + 
          8 S^2 
          d u^2
     \right] 
\end{equation} 
where upon applying Eq. \eqref{connection} we will have 
the desired Ruppeiner metric from Eq. \eqref{rnwein} \cite{abp1}: 
\begin{equation} \label{rnrupp} 
     d s^2_R = 
     \frac{1}{T} 
     \left( d s^2_W \right) = 
     - \frac{1}{2 S} d S^2 + 
     \frac{4 S}{1 - u^2} d u^2, 
\qquad  \text{where} \quad 
     T = 
     \frac{\partial M}{\partial S} = 
     \frac{1}{4 \sqrt{S}} 
     \left( 1 - u^2 \right). 
\end{equation} 
Using Eq. \eqref{rnentropy} and Eq. \eqref{rnrupp}, 
we can write the variable map 
$\eta_i = \eta_i (\bm{\widetilde{\eta}})$ as: 
\begin{equation} \label{varmap1} 
     \begin{split} 
          \eta_1 &= 
          M = 
          \frac{\sqrt{S}}2 
          \left( 
               1 + u^2 
          \right) 
     \\ 
          \eta_2 &= 
          Q = 
          u \sqrt{S}, 
     \end{split} 
\quad \Rightarrow \quad 
     \frac{\partial \eta_i}{\partial \widetilde \eta_a} = 
     \left( 
          \begin{array}{cc}
               \dfrac1{2 \sqrt{S}} 
               \left( 1 + u^2 \right) 
               & 
               \sqrt{S} u 
          \\ 
               \dfrac{u}{2 \sqrt{S}} 
               & 
               \sqrt{S}
          \end{array} 
     \right) 
\end{equation} 
Using the entropy Eq. \eqref{rnentropy} as the potential, 
we can write according to  Eq. \eqref{gradsol} and Eq. \eqref{dualcoor}: 
\begin{equation} \label{rndualcoor1} 
     \begin{split} 
          \theta^M &= 
          \frac{\partial S}{\partial M} = 
          2 \left( 
               2 M + 
               \frac{2 M^2 - Q^2}{\sqrt{M^2 - Q^2}}     
          \right) = 
          \frac{4}{1 - u^2} \sqrt{S} = 
          C \text{e}^{- t} 
     \\ 
          \theta^Q &= 
          \frac{\partial S}{\partial Q} = 
          - 2 Q 
          \left( 
               1 + 
               \frac{M}{\sqrt{M^2 - Q^2}}
          \right) = 
          - \frac{4 u}{1 - u^2} \sqrt{S} = 
          D \text{e}^{- t}. 
     \end{split}      
\end{equation} 
where $C, D$ are constants. 
From Eq. \eqref{rndualcoor1}, we can see that 
\begin{equation} \label{sols1} 
     u = 
     - \frac{\theta^Q}{\theta^M} = 
     - \frac{D}{C}, 
\quad 
     S = 
     \left( 
          \frac{C^2 - D^2}{4 C} 
     \right)^2 
     \text{e}^{- 2t}, 
\quad 
     Q = 
     \pm \frac{D}{C} \sqrt{S} = 
     \pm \frac{D}4 
     \left( 
          \frac{D^2}{C^2} - 1 
     \right) \text{e}^{- t}.
\end{equation} 
The value of $C$ can be determined using Eq. \eqref{varmap1} 
and Eq. \eqref{rndualcoor1} to write according to Eq. \eqref{transform}: 
\begin{equation} \label{rndualcoor2} 
     \left( 
          \begin{array}{c}
               \widetilde \theta^S 
          \\ 
               \widetilde \theta^u
          \end{array} 
     \right) = 
     \left( 
          \frac{\partial \eta_i}{\partial \widetilde \eta_a} 
     \right)^t 
     \left( 
          \begin{array}{c}
               \theta^M 
          \\ 
               \theta^Q
          \end{array} 
     \right) = 
     \left( 
          \begin{array}{cc}
               \dfrac{1 + u^2}{2 \sqrt{S}} 
               & 
               \dfrac{u}{2 \sqrt{S}} 
          \\ 
          &
          \\
               \sqrt{S} u 
               & 
               \sqrt{S}
          \end{array} 
     \right) 
     \dfrac{4 \sqrt{S}}{1 - u^2} 
     \left( 
          \begin{array}{c}
               1 
          \\ 
               - u 
          \end{array} 
     \right) = 
     \left( 
          \begin{array}{c}
               \dfrac{2}{1 - u^2} 
          \\ 
               0
          \end{array} 
     \right)    
\end{equation} 
\begin{equation} \label{constant} 
     \widetilde \theta^u = 0 
\quad \Rightarrow \quad 
     u = 0 
\quad , \quad 
     \widetilde \theta^S = 2. 
\end{equation}

\section{Conclusion and Discussion} 

We demonstrated how Hamilton-Jacobi equations 
for Riemannian metrics are similar to the gradient 
flow equations in IG, thus showing that such gradient 
flow equations describe Riemannian metrics. 
We also showed that such Riemannian metrics 
derived from gradient-flow equations have a conformal 
factor, and thus, resemble optical metrics. 
Then we showed that the two types of gradient flow 
equations are related to each other via canonical 
transformation, and thus are dual to each other. 

Next, we discussed different deformations of IG related 
to the gradient-flow equations. 
We showed that deforming the metric does not alter 
the dynamics, but deforming the gradient-flow equations 
will lead to distorted gradient-flow equations and RF metrics. 

We then tested our theories on the Gaussian model. 
From the Riemannian metric we described a spacetime metric, 
then showed that mechanics derived from the constraint were 
valid and consistent in supporting the gradient-flow equations. 
This consistency was shown to continue even for a RF metric 
described in the Gaussian setting, by supporting the deformed 
gradient-flow equations.

Finally, we applied our theory to the study of geometric 
thermodynamics of black holes. 
We developed upon previous work by Aman, Bengtsson, 
and Pidokrajt on flat metrics of IG in black hole thermodynamics 
by regarding them as to be dynamically evolving systems, 
and describing the dynamical evolution of the thermodynamic 
variables.

\section*{Acknowledgments}

We wish to acknowledge P. Guha for fruitful discussions that 
led to the production of this article.

\end{document}